\documentclass[onecolumn,sort&compress,numbers]{els-mrw} 
\usepackage{bm}
\usepackage{braket}
\usepackage{amsmath}
\usepackage{amssymb}
\usepackage{amsfonts}
\usepackage{amsthm}
\usepackage{makeidx}
\usepackage{mathrsfs}
\usepackage{graphicx}
\usepackage{booktabs}
\usepackage{array}
\usepackage{placeins}
\usepackage{txfonts}
\usepackage{helvet}
\usepackage{ulem}
\usepackage{lineno}

\usepackage{tikz, tikz-feynhand}
\usetikzlibrary{intersections, calc, arrows.meta, positioning}
\usetikzlibrary{fadings, patterns}
\tikzfading[name=fade top,
            top color=transparent!0,
            bottom color=transparent!100]
\tikzfading[name=fade out,
            inner color=transparent!20,
            outer color=transparent!80]
            
\usepackage[colorlinks=true,pdfstartview=Fit,linkcolor=blue,citecolor=magenta,urlcolor=blue,bookmarks=true,bookmarksnumbered=true]{hyperref}

\providecommand{\doi}[1]{}
\renewcommand{\doi}[1]{\href{https://doi.org/#1}{doi:#1}}
\providecommand{\eprint}[2][]{}
\renewcommand{\eprint}[2][]{\href{https://arxiv.org/abs/#2}{\texttt{#2}}}

\begin{document}

\chapter[Axial $U(1)$ Symmetry Breaking in Hot QCD]{Axial $U(1)$ Symmetry Breaking in Hot QCD: From Topology to the Chiral Phase Transition}

\author[1]{Heng-Tong Ding}
\address[1]{Key Laboratory of Quark \& Lepton Physics (MOE) and Institute of Particle Physics, Central China Normal University, Wuhan 430079, China}

\maketitle
\begin{abstract}[Abstract]
Heating matter can restore symmetries spontaneously broken at low temperature.
The axial $U(1)$ symmetry of quantum chromodynamics (QCD) is different:
it is present in the classical theory with massless quarks but is broken
upon quantization by the axial anomaly. The anomaly persists at every
temperature, yet its observable effects can weaken. Can hot matter
nevertheless behave as though this symmetry were restored at long distances?
Whether such effective axial $U(1)$ restoration occurs depends on how
microscopic quark and gluon dynamics governs the strength and spatial
range of axial breaking.
	
This review brings together theoretical developments and first-principles
lattice QCD calculations. We examine how gluon-field topology shapes the
low-lying Dirac modes and their correlations, how these modes contribute
to axial breaking at different spatial scales, and what this implies for the order and critical behavior of the chiral phase transition 
as quark masses approach zero.
\end{abstract}

\section{Introduction}

Heating matter can restore a symmetry broken by its low-temperature state.
But what happens when a symmetry is broken by quantum effects themselves?
Quantum chromodynamics (QCD), the theory of quarks and gluons, provides a
concrete setting for this question. Its classical flavor-singlet axial
symmetry, denoted $U_A(1)$, is broken by the \emph{axial anomaly}. The
anomaly remains present at every temperature, yet its influence on the
collective behavior of quarks and gluons can change. This review asks
whether, and through which microscopic mechanisms, that influence remains
important at long distances near the QCD chiral phase transition.

For $N_f$ massless quark flavors, or species, the classical QCD Lagrangian
allows independent rotations of the left- and right-handed quark fields.
These two components are distinguished by their \emph{chirality}. A
$U_A(1)$ transformation rotates them by opposite phases, with the same
transformation applied to every flavor; it is therefore called a flavor-singlet axial transformation. 
Classically, the associated axial current is conserved.
Quantization changes its conservation law, producing the axial
anomaly~\cite{Adler:1969gk,Bell:1969ts,Fujikawa:1979ay,Fujikawa:1980eg}.
This anomalous breaking is distinct from \emph{explicit} breaking by nonzero quark masses and from \emph{spontaneous} breaking, 
in which the vacuum or equilibrium state does not share a symmetry of its equations.

The anomaly links quark chirality to \emph{gauge-field topology}: global
properties of gluon configurations characterized by a topological charge.
Through the index theorem, this charge fixes an imbalance between left-
and right-handed zero-eigenvalue modes of the Dirac operator, the operator
governing quark propagation in a gluon background~\cite{Atiyah:1967ih,tHooft:1976snw}.
The consequences are visible in the hadron spectrum. Most famously, the
anomaly helps explain why the $\eta'$ meson is much heavier than the light
mesons associated with spontaneous chiral symmetry breaking. The
Witten--Veneziano relation makes the connection to topology
explicit~\cite{Witten:1979vv,Veneziano:1979ec}. Topologically nontrivial
fields also induce the multi-fermion 't Hooft interaction, which couples
several quark fields and distinguishes processes related by
$U_A(1)$~\cite{tHooft:1976snw}.

At low temperature, a nonzero light-quark condensate---an expectation
value of a quark--antiquark operator---signals spontaneous breaking of
non-singlet chiral symmetry in the massless theory. For physical up, down,
and strange quark masses and zero net baryon density, heating produces a
smooth \emph{chiral crossover} near the pseudocritical temperature
$T_{\rm pc}\simeq155$--$160$ MeV
~\cite{Aoki:2006we,HotQCD:2018pds,Borsanyi:2020fev,RubenKara:2024krv}. The condensate decreases
rapidly, while meson correlations, topological fluctuations, and the
distribution of Dirac eigenvalues near zero reorganize. 
A natural question then arises: as heating weakens non-singlet chiral
symmetry breaking, does it also suppress the long-distance effects of the
anomaly?

The answer to this question matters especially for the nature of the chiral phase transition.
A smooth crossover at physical quark masses does not itself define an
exact critical universality class. The sharper theoretical question
concerns the \emph{two-flavor chiral limit}, in which the up and down quark
masses vanish while the strange-quark mass is held fixed. In this limit,
anomaly-induced interactions can influence whether the transition is
first order or continuous and, if continuous, its \emph{universality
	class}: the common scaling behavior shared by systems with different
microscopic details~\cite{Pisarski:1983ms,Pelissetto:2013hqa,
	Pisarski:2024esv}. Hot QCD thus connects the quantum physics of quarks and
gluons to the statistical physics of collective fluctuations.
The review also considers how this question and the relevant diagnostics
change when three or more quark flavors become massless.

However, heating does not remove the anomaly itself: the anomalous
\emph{Ward identity} remains intact, even though the anomaly's infrared
manifestations can weaken or vanish in particular observables. The
vanishing of these infrared effects after the relevant limits are taken
is referred to here as \emph{effective $U_A(1)$ restoration}. Such
restoration can be tested by comparing meson correlation functions for
operators related by $U_A(1)$. These functions describe how
quark--antiquark fluctuations at separated points are related; their
spacetime integrals are called \emph{susceptibilities}. Agreement of these
correlation functions or susceptibilities is not implied by restoration
of the non-singlet chiral symmetry of the two light flavors,
$SU(2)_L\times SU(2)_R$~\cite{Aoki:2012yj,Buchoff:2013nra,
	Tomiya:2016jwr,Ding:2020xlj,Kaczmarek:2021ser}.
Such agreement, when found, applies to the observables tested; it does not
establish the symmetry of all correlation functions. To investigate the
microscopic origin of these signals, one can relate susceptibility
differences between $U_A(1)$ partners to the eigenvalues of the Dirac
operator, collectively called the Dirac spectrum~\cite{Chandrasekharan:1995gt}.

The Dirac spectrum, in turn, connects these susceptibility tests to
statistical and condensed-matter physics. Correlations among small Dirac
eigenvalues encode
fluctuations of the chiral condensate, linking microscopic spectral
structure to collective behavior near the
transition~\cite{Ding:2023oxy}. Dirac modes can also become spatially
localized, a phenomenon related to Anderson localization of electron wave
functions in disordered materials~\cite{Anderson:1958vr,Garcia-Garcia:2006vlk}. Studying both
eigenvalue correlations and the spatial extent of the modes can help explain
how microscopic quark dynamics shapes meson correlations over different
distances.

The subject also reaches beyond the equilibrium chiral transition. Hot
quark--gluon matter was present in the early universe and is recreated
briefly in relativistic heavy-ion collisions. Through the anomaly,
transitions between gluon configurations of different topology can generate
an imbalance between left- and right-handed quarks. In a magnetic field,
this imbalance can drive an electric current along the field---the
\emph{chiral magnetic
	effect}~\cite{Fukushima:2008xe,Kharzeev:2024zzm}. Equilibrium fluctuations
of the total topological charge, measured by the topological susceptibility,
also determine the temperature-dependent QCD-induced mass of the axion, a
proposed dark-matter particle~\cite{Borsanyi:2016ksw,Petreczky:2016vrs}. This cosmological
application and effective axial restoration are related, but they are not
the same question.

Near the crossover, the equilibrium problem is \emph{nonperturbative}:
the QCD coupling is not small enough for a reliable perturbative
expansion. Lattice QCD provides
a first-principles treatment by evaluating the equilibrium theory
numerically on a spacetime grid without this expansion.
The calculations reviewed here find that many $U_A(1)$-breaking signals
weaken with increasing temperature. The unresolved issue is whether the
relevant signals vanish in the light-quark chiral limit near the
transition, and what microscopic dynamics controls their
behavior~\cite{Tomiya:2016jwr,Ding:2020xlj,Aoki:2020noz,Dick:2015twa,Kaczmarek:2021ser,
	Kaczmarek:2023bxb,Kovacs:2023vzi,JLQCD:2024xey,Alexandru:2024tel,Gavai:2024mcj,Aoki:2025mue}.

Connecting these calculations to the chiral-limit question requires
controlling their dependence on quark masses, simulation volume, and
lattice spacing. The corresponding limits are the chiral (vanishing
light-quark masses), thermodynamic (infinite-volume), and continuum
(vanishing lattice-spacing) limits, whose order can matter. Throughout
the review, conclusions are therefore tied to a specified observable,
temperature, quark masses, and
limiting procedure. Different ways of representing quarks on the lattice
allow results to be compared across formulations with different systematic
uncertainties.

The opportunity extends beyond determining whether effective $U_A(1)$
restoration occurs to explaining the microscopic dynamics behind the
temperature dependence of axial-breaking effects. How does heating reorganize the topological
fluctuations of gluon fields, and how is this reorganization reflected
in quark propagation and meson correlations? Why can the effects of
the anomaly remain prominent in some observables while becoming weak
in others? Which anomaly-induced interactions remain important as
progressively longer distances are probed? Connecting topological
observables, the Dirac spectrum, and correlation functions with
microscopic models and effective theories offers ways
to investigate these questions. The central goal is to understand how
an exact quantum anomaly shapes the changing correlations and
collective behavior of hot QCD.

This perspective guides the organization of the chapter.
Sections~\ref{sec:basics},\ref{sec:nonzeroT},\ref{sec:history} introduce the anomaly and its connections to topology
and the hadron spectrum, explain the finite-temperature transition
problem, and trace the historical development of the subject.
Sections~\ref{sec:diagnostics} and~\ref{sec:lattice-systematics} develop the principal diagnostics and lattice
methods, including the systematic uncertainties relevant to their
interpretation. Section~\ref{sec:lattice-evidence} reviews the numerical
evidence in three connected themes: topology and the infrared spectrum,
meson correlations and the spatial structure of Dirac modes, and the
approach to the chiral phase transition as quark masses and flavor content change.
Section~\ref{sec:questions} follows the same sequence to assess microscopic
mechanisms, the spatial range of axial breaking, and its relation to critical
behavior, identifying focused directions for further investigation.
Section~\ref{sec:perspective} closes with the broader connections.

\section{The axial anomaly in QCD: a pedagogical overview}
\label{sec:basics}

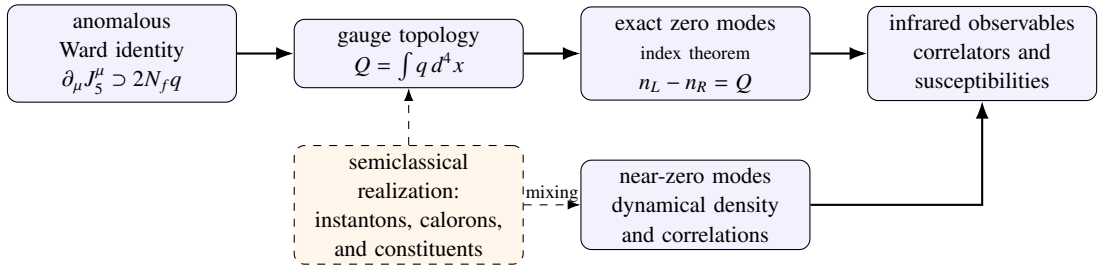
\begin{figure}[htbp]
\centering
\begin{tikzpicture}[
  node distance=0.75cm,
  box/.style={draw,rounded corners,align=center,minimum height=0.9cm,
              text width=0.17\textwidth,fill=blue!5},
  every edge/.style={draw,-{Latex},thick}]
\node[box] (ward) {anomalous Ward identity\\$\partial_\mu J_5^\mu\supset2N_fq$};
\node[box,right=of ward] (top) {gauge topology\\$Q=\int q\,d^4x$};
\node[box,right=of top] (modes) {exact zero modes\\{\footnotesize index theorem}\\$n_L-n_R=Q$};
\node[box,right=of modes] (obs) {infrared observables\\correlators and\\susceptibilities};
\path (ward) edge (top) (top) edge (modes) (modes) edge (obs);
\node[draw,dashed,rounded corners,below=0.75cm of top,align=center,
      text width=0.17\textwidth,fill=orange!8] (semi)
      {semiclassical realization:\\instantons, calorons, and constituents};
\node[box] (near) at (modes |- semi)
      {near-zero modes\\dynamical density\\and correlations};
\draw[dashed,-{Latex}] (semi) -- (top);
\draw[dashed,-{Latex}] (semi) --
      node[above,inner sep=1pt,font=\scriptsize] {mixing} (near);
\draw[thick,-{Latex}] (near.east) -| (obs.south);
\end{tikzpicture}
\caption{Connections between the exact axial anomaly and finite-temperature
diagnostics.  The index theorem fixes the net chirality of exact zero modes.
Near-zero modes and their correlations depend on dynamics and can occur even
at $Q=0$.  Dashed links illustrate semiclassical realizations; the anomaly
itself does not rely on a dilute-instanton description.  The strength of the
infrared signal is dynamical.}
\label{fig:anomaly-chain}
\end{figure}

The finite-temperature problem is easiest to understand after separating four
ideas that are sometimes compressed into the word ``chiral'': handedness,
flavor symmetry, spontaneous symmetry breaking, and the quantum anomaly.  This
section introduces each idea and then assembles the chain of relations shown
schematically in Fig.~\ref{fig:anomaly-chain}.

\subsection{Classical chiral symmetry and the quantum anomaly}

For $N_f$ quark flavors the QCD Lagrangian is
\begin{equation}
 \mathcal{L}_{\mathrm{QCD}}=
 -\frac{1}{4}F_{\mu\nu}^{a}F^{a\,\mu\nu}
 +\bar\psi\,(i\gamma^\mu D_\mu-M)\psi ,
 \label{eq:qcd-lagrangian}
\end{equation}
where $\psi=(u,d,\ldots)^T$ is a vector in flavor space and $M$ is the
quark-mass matrix.  A massless fermion can be separated into components of
definite chirality,
\begin{equation}
 \psi_L=P_L\psi,\qquad \psi_R=P_R\psi,\qquad
 P_{L,R}=\frac{1\mp\gamma_5}{2}.
\end{equation}
For an ultrarelativistic particle, chirality is closely related to the more
intuitive notion of helicity, the projection of spin along momentum.  Chirality
is the useful field-theory concept because the QCD interaction with gluons does
not mix $\psi_L$ and $\psi_R$ when $M=0$.

The left- and right-handed fields can therefore be rotated independently,
\begin{equation}
 \psi_L\longrightarrow L\psi_L,
 \qquad
 \psi_R\longrightarrow R\psi_R,
 \qquad L,R\in U(N_f).
\end{equation}
Ignoring discrete identifications among group centers, the classical symmetry
is conventionally displayed as
\begin{equation}
 U(N_f)_L\times U(N_f)_R
 \simeq SU(N_f)_L\times SU(N_f)_R\times U_V(1)\times U_A(1).
 \label{eq:classical-symmetry}
\end{equation}
The vector factor $U_V(1)$ rotates left- and right-handed fields by the same
phase and expresses quark-number conservation (or baryon-number conservation
after a change of normalization).  A singlet axial rotation instead acts with
opposite phases,
\begin{equation}
 \psi\longrightarrow e^{i\alpha\gamma_5}\psi,
 \qquad
 \psi_L\longrightarrow e^{-i\alpha}\psi_L,
 \qquad
 \psi_R\longrightarrow e^{i\alpha}\psi_R.
 \label{eq:ua-transformation}
\end{equation}
The last two relations follow from the decomposition
$\psi=\psi_L+\psi_R$, together with $\gamma_5\psi_L=-\psi_L$ and
$\gamma_5\psi_R=\psi_R$.
The adjective \emph{singlet} means that every flavor receives the same phase.

A mass term connects left- and right-handed fields,
$\bar\psi_L M\psi_R+\bar\psi_R M^\dagger\psi_L$, and is not invariant under
independent rotations.  Nonzero quark masses therefore break both the
non-singlet axial transformations and Eq.~(\ref{eq:ua-transformation})
explicitly.  The anomaly is a separate source of breaking that remains even
when all entries of $M$ are zero.

Noether's theorem associates the singlet axial transformation with
\begin{equation}
 J_5^\mu=\sum_{f=1}^{N_f}\bar\psi_f\gamma^\mu\gamma_5\psi_f.
\end{equation}
The classical equations of motion give
\begin{equation}
 \partial_\mu J_5^\mu
 =2i\sum_{f=1}^{N_f}m_f\bar\psi_f\gamma_5\psi_f,
 \label{eq:classical-ward}
\end{equation}
which vanishes in the massless limit.  In the quantum theory, however, the
current and its products require ultraviolet regularization.  A regulator can
preserve gauge invariance or the singlet axial conservation law, but not both.
Gauge invariance is indispensable for a consistent gauge theory, and the
renormalized identity becomes~\cite{Adler:1969gk,Bell:1969ts}
\begin{equation}
 \partial_\mu J_5^\mu
 =2i\sum_{f=1}^{N_f}m_f\bar\psi_f\gamma_5\psi_f+2N_f q(x),
 \qquad
 q(x)=\frac{g^2}{32\pi^2}F_{\mu\nu}^a\widetilde F^{a\,\mu\nu},
 \label{eq:anomalousward}
\end{equation}
with
$\widetilde F^{a\,\mu\nu}=\tfrac12\epsilon^{\mu\nu\rho\sigma}
F^a_{\rho\sigma}$.  The second term is the axial anomaly.  Its coefficient is
protected from higher-order perturbative corrections by the Adler--Bardeen
theorem; Bardeen's analysis gives the corresponding gauge-invariant framework
for anomalous Ward identities~\cite{Adler:1969er,
Bardeen:1969md}.  Increasing the temperature does not remove it.  An
explicit finite-temperature analysis by Itoyama and Mueller reached the same
conclusion: the thermal medium changes the allowed tensor structures of
correlation functions, but not the operator anomaly
~\cite{Itoyama:1982up}.  This is the historical origin of the important
distinction between the anomaly and the size of its infrared manifestations.

Two complementary derivations illuminate different aspects of
Eq.~(\ref{eq:anomalousward}).  In perturbation theory, the anomalous term is
exposed by the triangle graph containing one axial-current insertion and two
gauge vertices~\cite{Adler:1969gk,Bell:1969ts}.  In Fujikawa's path-integral
derivation, the classical action at $M=0$ is invariant under an axial rotation,
but the infinite-dimensional fermion integration measure is not
~\cite{Fujikawa:1979ay,Fujikawa:1980eg}.  Regulating its Jacobian with
eigenfunctions of the Dirac operator produces precisely the density $q(x)$.
Fujikawa's derivation makes especially clear how ultraviolet regularization
produces the topological density and, through the index theorem, connects the
integrated anomaly to chiral zero modes of the Dirac operator.

The integrated density
\begin{equation}
 Q=\int d^4x\,q(x)
 \label{eq:topological-charge}
\end{equation}
is the topological charge of a sufficiently smooth Euclidean gauge field and
takes integer values.  Under the flavor-singlet axial rotation in
Eq.~(\ref{eq:ua-transformation}), the classical massless action is invariant,
but the fermion measure acquires an anomalous phase.  Thus continuous
$U_A(1)$ is not an exact symmetry of quantum QCD.  The phase is unity for
every integer topological charge $Q$ when $\alpha=k\pi/N_f$, with integer
$k$, leaving a discrete $Z_{2N_f}$ subgroup unbroken by the anomaly.
Equivalently, an axial change of variables shifts the coefficient $\theta$
of a possible term $\theta q(x)$ in the QCD
Lagrangian.  A non-anomalous quantum symmetry could not change a parameter of
the theory in this way; the shift of $\theta$ is therefore the signature of the
$U_A(1)$ anomaly.

\subsection{Topology, instantons, and the Dirac spectrum}

In Euclidean spacetime the massless Dirac operator $D\!\!\!/$ anticommutes with
$\gamma_5$.  If a nonzero mode satisfies
\begin{equation}
 D\!\!\!/\,\psi_\lambda=i\lambda\psi_\lambda,
\end{equation}
then $\gamma_5\psi_\lambda$ has eigenvalue $-i\lambda$.  Nonzero eigenvalues
therefore occur in pairs.  A zero mode need not have a partner and can be
chosen to be purely left- or right-handed.  The Atiyah--Singer index theorem
relates this spectral imbalance to topology,
\begin{equation}
 \mathrm{index}(D\!\!\!/)=n_L-n_R=Q,
 \label{eq:index-theorem}
\end{equation}
up to the overall sign convention for $Q$~\cite{Atiyah:1967ih}.
Here $n_L$ and $n_R$ count the linearly independent left- and right-handed
zero modes of the massless Dirac operator in a given gauge field.  Its
\emph{index} is their difference, $n_L-n_R$, measuring the net chirality of
the zero modes rather than their total number.
Thus gluon topology leaves a countable fermionic imprint: a configuration with
$Q\ne0$ has unpaired chiral zero modes.

Exact zero modes and \emph{near-zero} modes must be distinguished.  The index
theorem fixes only the difference $n_L-n_R$ and says nothing by itself about a
dense population at small nonzero $|\lambda|$.  Near-zero modes can arise when
topological objects of opposite charge interact, and they can remain numerous
even when the net topological charge is zero.  This distinction is important
in a finite volume.  Exact zero modes can strongly affect observables at small
quark mass, although their density may vanish as the volume tends to infinity.
By contrast, near-zero modes whose number grows with the volume can remain
important in that limit.

The Belavin--Polyakov--Schwartz--Tyupkin (BPST) instanton is the prototype localized, finite-action solution of the
four-dimensional Euclidean $SU(2)$ Yang--Mills equations and can be embedded in the $SU(3)$ gauge theory of QCD.  It
carries one unit of topological charge; an anti-instanton carries
the opposite charge~\cite{Belavin:1975fg}.  The index theorem
assigns a chiral zero mode to each light flavor in such a background.
Integrating over the fermions converts these zero modes into an effective
$2N_f$-fermion interaction, schematically
\begin{equation}
 \mathcal{L}_{\mathrm{'t\,Hooft}}
 \sim \det_{f,f'}(\bar\psi_R^f\psi_L^{f'})+\mathrm{h.c.}
 \label{eq:thooft-vertex}
\end{equation}
~\cite{tHooft:1976rip,tHooft:1976snw}.  The determinant is invariant under
$SU(N_f)_L\times SU(N_f)_R$ but acquires a phase under $U_A(1)$.  Each term
in the flavor determinant contains $N_f$ bilinears and hence $2N_f$ fermion
fields, giving a four-fermion interaction for two flavors and a six-fermion
interaction for three.  Equation~(\ref{eq:thooft-vertex}) gives an intuitive microscopic
picture of how topology distinguishes channels related by a singlet axial
rotation.

Instantons should not be equated with the anomaly.  The anomalous Ward identity
is exact, whereas a gas of weakly interacting instantons is an approximation
whose accuracy depends on temperature and scale.  Near the chiral crossover,
gauge fields are strongly coupled and may involve interacting instantons and
anti-instantons, finite-temperature calorons, their monopole-like
constituents~\cite{Harrington:1978ve,Kraan:1998pm,
Lee:1998bb}, or fluctuations that admit no simple semiclassical label.
At asymptotically high temperature, by contrast, screening suppresses large
instantons and a dilute instanton gas becomes better motivated.
First-principles lattice QCD calculations are needed to determine whether
this description is quantitatively reliable at a given temperature.

\subsection{Chiral symmetry breaking and zero-temperature consequences of the $U_A(1)$ anomaly}

At low temperature the QCD vacuum spontaneously selects an orientation in
chiral space,
\begin{equation}
 SU(N_f)_L\times SU(N_f)_R\longrightarrow SU(N_f)_V.
\end{equation}
The chiral condensate is an order parameter in the massless theory.  For one
flavor $f$, with the conventional negative sign for
$\langle\bar\psi_f\psi_f\rangle$, the Banks--Casher relation connects it to
the spectral density of the Dirac operator,
\begin{equation}
 \lim_{m\to0}\lim_{V\to\infty}
 \langle\bar\psi_f\psi_f\rangle=-\pi\rho(0),
 \label{eq:banks-casher}
\end{equation}
where $\rho(\lambda)$ counts eigenvalues of a single-flavor Dirac operator
per unit four-volume, and $\rho(0)$ denotes its limiting density at the
origin~\cite{Banks:1979yr}.  An accumulation of near-zero modes is thus
the spectral signature of spontaneous chiral symmetry breaking.  The order of
limits in Eq.~(\ref{eq:banks-casher}) matters: no continuous symmetry breaks
spontaneously in a finite volume.

$U_A(1)$ is different because it is not an exact continuous symmetry of
quantum QCD to begin with.  Its finite-temperature ``restoration'' can only
mean that anomaly-induced effects disappear from a specified infrared sector.
Table~\ref{tab:symmetry-mechanisms} summarizes the distinction that will recur
throughout the review.

\begin{table}[htbp]
\centering
\begin{tabular}{p{0.17\textwidth}p{0.25\textwidth}p{0.25\textwidth}p{0.22\textwidth}}
\toprule
Symmetry & Status in massless quantum QCD & Low-temperature realization & Finite-temperature question \\
\midrule
$U_V(1)$ & Exact (apart from electroweak baryon-number effects outside pure QCD) & Unbroken & Remains exact and unbroken; no restoration question arises. \\
$SU(N_f)_L\times SU(N_f)_R$ & Exact for $M=0$ & Spontaneously broken to $SU(N_f)_V$ & Does the condensate vanish and do non-singlet chiral partners become degenerate? \\
$U_A(1)$ & Broken by the quantum anomaly even for $M=0$ & Anomaly effects are large in the infrared & Do specified long-distance $U_A(1)$-breaking observables vanish in the relevant limits? \\
\bottomrule
\end{tabular}
\caption{Symmetries and their realization in QCD with $N_f$ massless quark flavors at zero chemical potentials. ``Restoration'' has a
literal thermodynamic meaning for spontaneous breaking but only an effective,
observable-dependent meaning for $U_A(1)$.}
\label{tab:symmetry-mechanisms}
\end{table}

The best-known consequence is the $U(1)$ problem.  If the full classical
$U(3)_L\times U(3)_R$ symmetry were spontaneously broken to $U(3)_V$, one
would expect nine light pseudoscalar modes.  The octet associated with the
non-singlet axial generators is recognizable in the light pseudoscalar mesons,
but the predominantly singlet $\eta'$ is much heavier.  Because the anomalous
$U_A(1)$ is not a symmetry, Goldstone's theorem never requires a ninth light
state.

At leading order in the large-number-of-colors expansion, the
Witten--Veneziano relation sharpens this statement,
\begin{equation}
 m_{\eta'}^2+m_\eta^2-2m_K^2
 \simeq \frac{2N_f}{f_\pi^2}\,\chi_{\mathrm{YM}},
 \label{eq:witten-veneziano}
\end{equation}
where $\chi_{\mathrm{YM}}$ is the topological susceptibility of pure
Yang--Mills theory and normalization-dependent corrections are understood
~\cite{Witten:1979vv,Veneziano:1979ec}.  The unusually large singlet mass is
thereby tied to fluctuations of topological charge~\cite{Zahed:2026wag}.

Anomalies also control observable decay amplitudes.  The classic example
$\pi^0\to\gamma\gamma$ originates from the electromagnetic anomaly of a
flavor-nonsinglet axial current~\cite{Adler:1969gk,Bell:1969ts}.  It is
distinct from the flavor-singlet $U_A(1)$ anomaly generated by QCD gauge
fields, which is the subject of this review.  Nevertheless, it provides a
familiar example of how quantum effects modify a classical current-conservation
law and produce an observable consequence.  The gluonic anomaly likewise
affects the $\eta$--$\eta'$ system, scalar and pseudoscalar interactions, the
$\theta$ dependence of the vacuum, and the QCD input to axion physics.

These zero-temperature facts establish the baseline.  The exact relation
between chirality and topology remains true after heating, but the thermal
ensemble can change the abundance and correlations of topological gauge
fields and the low-lying Dirac modes associated with them.  The fate of
$U_A(1)$ at nonzero temperature is the question of how much of this
anomaly-driven infrared structure survives.

\section{Why axial symmetry breaking matters at nonzero temperature}
\label{sec:nonzeroT}

Finite temperature changes the probability with which QCD samples different
gauge fields; it does not change the operator identity
Eq.~(\ref{eq:anomalousward}).  The problem is therefore dynamical: do
topological gauge-field fluctuations and the associated near-zero Dirac modes
continue to affect correlations between widely separated spatial points near
the thermal transition?  Such effects can appear in axial-partner correlation
lengths and, at a continuous chiral phase transition, in the critical behavior.

Three questions organize the discussion.  We first clarify what can be meant
by effective restoration at nonzero quark mass and in the chiral limit.  We
then ask how the infrared strength of axial breaking can affect the order and
universality of the transition, including its flavor dependence.  
Finally, we relate topology to axial observables and axion physics, distinguish the crossover from the high-temperature regime, and identify criteria for comparing calculations.

\subsection{Thermal QCD and effective restoration}

An equilibrium quantum system at temperature $T$ has partition function
$Z=\mathrm{Tr}\,e^{-H/T}$.  In the Euclidean path integral this becomes a
theory with compact time direction of length $1/T$; gluon fields are periodic
and quark fields antiperiodic around that direction.  At low temperature, QCD
is well described in terms of hadrons and the light-quark chiral condensate is
large.  At high temperature, quark and gluon degrees of freedom become the
more economical description and the condensate is strongly reduced~\cite{Borsanyi:2025ttb}.

With physical up, down, and strange masses, thermodynamic quantities change
smoothly rather than singularly.  Different crossover observables need not
peak or turn over at exactly the same temperature, but modern
continuum-extrapolated lattice-QCD calculations place the chiral
pseudocritical region near $155$--$160$ MeV
~\cite{HotQCD:2018pds,Borsanyi:2020fev,RubenKara:2024krv}.  We denote a
pseudocritical temperature by $T_{\rm pc}$.  The symbol $T_c$ will be reserved
for a genuine critical temperature in a massless limit, or used when following
the notation of a cited calculation.

The absence of a singularity at physical masses does not remove the underlying
symmetry question.  The up and down quarks are light enough that the system
shows approximate scaling associated with chiral criticality.  One can also
simulate several light-quark masses and extrapolate toward $m_l=0$.  It is in
this two-flavor chiral limit, usually at a fixed physical strange-quark mass,
that the order and universality class of the transition become sharply
defined.

Several statements of increasing strength are often called ``$U_A(1)$
restoration.''  They should not be conflated.

\begin{enumerate}
 \item A particular difference of two-point functions or integrated
 susceptibilities can become numerically small at nonzero quark mass.
 \item That difference can vanish after the thermodynamic and chiral limits.
 This establishes effective restoration in the corresponding two-point
 sector.
 \item All infrared correlation functions can become invariant under axial
 rotations.  This is a stronger statement; equality of one pair of
 susceptibilities is necessary but not sufficient.
\end{enumerate}

The first statement is what a finite simulation measures directly.  Reaching
the second requires controlled thermodynamic and chiral extrapolations.  The
third may require higher-point functions and assumptions about analyticity of
the free energy or the Dirac spectral density~\cite{Aoki:2012yj}.  Throughout
this review, ``effective restoration'' refers to one of these first three
meanings and will be qualified by the observable and limits being tested.

Topology and the Dirac spectrum are therefore diagnostics, not alternative
definitions of the symmetry. If any quark is exactly massless, the vacuum
energy becomes independent of $\theta$ and the topological susceptibility
vanishes, yet anomaly-induced multi-fermion correlations can remain nonzero.
For the Dirac spectrum, suppression of a spectral contribution does not
require a literal interval without eigenvalues: a sufficiently depleted
gapless density can make a specified spectral integral vanish.
Section~\ref{sec:diagnostics} makes these statements quantitative.
Before developing these diagnostics in detail, however, we first examine
how axial breaking enters the long-distance effective theory of the
chiral phase transition.

\subsection{The chiral phase transition, universality, and flavor dependence}
\label{sec:chiral-transition}

For the chiral phase transition, the physical question is whether anomalous
interactions remain important for meson fluctuations on the scales that
control critical behavior. Near a continuous transition, the spatial
correlation length $\xi$ of the chiral order parameter grows much larger
than the thermal length $1/T$. To examine the anomaly's role on this growing
scale, we need an effective theory of these collective fluctuations.

At finite temperature, bosonic Matsubara frequencies are
$\omega_n=2\pi nT$, while fermionic frequencies are
$\omega_n=(2n+1)\pi T$.  A nonzero Matsubara frequency introduces a thermal
scale in a spatial propagator.  When $\xi\gg1/T$, nonzero-frequency modes
can therefore be integrated out of the theory for static critical
fluctuations.  The elementary quark fields have no zero-frequency mode.
Under the usual assumption that only the chiral order-parameter fields
become critical, their static long-distance behavior can be described by a
three-dimensional Landau--Ginzburg--Wilson theory.

For QCD, the order-parameter field is a complex $N_f\times N_f$ matrix
representing a color-singlet quark bilinear,
\begin{equation}
 \Phi_{ij}\sim\bar\psi_{Rj}\psi_{Li},
 \qquad \Phi\longrightarrow L\Phi R^\dagger,
\end{equation}
Although its constituents are fermions, this bilinear is bosonic: the two
minus signs from the quarks' antiperiodic boundary conditions cancel, so it
is periodic and can have a static mode.  The expectation value
$\langle\Phi\rangle$ describes the chiral condensate, while fluctuations of
$\Phi$ describe collective scalar and pseudoscalar modes.  The quarks'
effects remain encoded in the interactions of this field.  The corresponding
schematic effective Lagrangian may be written as~\cite{Pisarski:1983ms}
\begin{align}
 \mathcal{L}_{\rm eff}={}&
 \mathrm{Tr}(\partial_k\Phi^\dagger\partial_k\Phi)
 +r\,\mathrm{Tr}(\Phi^\dagger\Phi)
 +u[\mathrm{Tr}(\Phi^\dagger\Phi)]^2 \nonumber\\
 &+v\,\mathrm{Tr}[(\Phi^\dagger\Phi)^2]
 -c_A(\det\Phi+\det\Phi^\dagger)+\cdots .
 \label{eq:lgw}
\end{align}
The determinant term represents $U_A(1)$ breaking.  Its coefficient $c_A$ is
an effective, temperature-dependent coupling, not the coefficient of the
operator anomaly in Eq.~(\ref{eq:anomalousward}).  At long distances, the
question is whether this and related interactions continue to affect axial-partner
correlations on the scale of the growing $\xi$, or become negligible in the
critical scaling limit.  This depends on their renormalization-group
flow as shorter-distance fluctuations are averaged out
~\cite{Pelissetto:2013hqa}.  A small signal in one integrated susceptibility
does not by itself settle this question: such an integral includes
contributions from short as well as long separations.
Section~\ref{sec:axial-range} returns to the distinction between the magnitude
of axial breaking and its spatial range.

Because $\det\Phi$ has degree $N_f$, the same anomalous interaction enters
different orders of the Landau expansion as the number of flavors changes.
It is therefore useful to complete the two-flavor universality argument before
turning to the distinct phase-transition and diagnostic questions for
$N_f\geq3$.

For two massless flavors, the determinant term is quadratic in $\Phi$ and can
directly separate the would-be axial partners.  If this breaking remains
substantial at the transition, the relevant chiral symmetry-breaking pattern is
\begin{equation}
 SU(2)_L\times SU(2)_R\longrightarrow SU(2)_V.
\end{equation}
This pattern has a direct realization in the light components of $\Phi$:
the scalar $\sigma\sim\bar\psi\psi$ and the three pion fields
$\pi^a\sim\bar\psi i\gamma_5\tau^a\psi$ form a four-component $O(4)$ vector
under chiral rotations.  A nonzero condensate selects the $\sigma$ direction,
while the unbroken $SU(2)_V$ rotates the three pion components among
themselves.  The local symmetry-breaking pattern is therefore
$O(4)\to O(3)$.  If the remaining
axial partners, the pseudoscalar singlet and scalar isovector fields, stay
massive at $T_c$, they can be integrated out of the critical theory, and a
continuous transition may belong to the
three-dimensional $O(4)$ universality class~\cite{Pisarski:1983ms}.

If $U_A(1)$ breaking is ineffective at the transition, the symmetry enlarges
to
\begin{equation}
 U(2)_L\times U(2)_R\longrightarrow U(2)_V.
\end{equation}
The original $4-\epsilon$ analysis did not find a stable fixed point and hence
suggested a first-order transition~\cite{Pisarski:1983ms}.  A subsequent
six-loop three-dimensional analysis reached the same conclusion for the
$U(2)\times U(2)$ theory~\cite{Butti:2003nu}.  Later high-order
analyses found evidence that a stable $U(2)\times U(2)$ fixed point can exist,
allowing a continuous transition in a universality class distinct from
$O(4)$~\cite{Pelissetto:2013hqa}.  Effective axial restoration therefore does
\emph{not} logically imply a first-order transition.  The dynamics still
matter, including whether the microscopic couplings lie in the region that
flows toward a stable fixed point.

\begin{table}[htbp]
\centering
\begin{tabular}{p{0.22\textwidth}p{0.30\textwidth}p{0.37\textwidth}}
\toprule
Infrared role of $U_A(1)$ & Symmetry-breaking pattern & Possible consequence for $m_{u,d}=0$ \\
\midrule
Substantial breaking & $SU(2)_L\times SU(2)_R\to SU(2)_V$ & Continuous $O(4)$ transition is allowed; small quark masses round it to a crossover. \\
Weak but nonzero breaking & Reduced chiral symmetry, with approximately enlarged symmetry on intermediate scales & Crossover between scaling regimes as axial breaking becomes relevant can produce sizable corrections to universal fits. \\
Effective absence & $U(2)_L\times U(2)_R\to U(2)_V$ & A distinct continuous transition or a first-order transition is possible, depending on the infrared flow. \\
\bottomrule
\end{tabular}
\caption{Common two-flavor scenarios at the chiral phase transition.  Substantial
axial breaking permits an $O(4)$ transition but does not guarantee continuity;
a first-order transition is also possible.  These are allowed infrared
scenarios, not a one-to-one inference from a single lattice observable.}
\label{tab:transition-scenarios}
\end{table}

Functional approaches also investigate these two-flavor possibilities~\cite{Fischer:2026vkc}.
The functional renormalization group finds a candidate
$U(2)\times U(2)$ fixed point~\cite{Grahl:2014fna}, while an analysis retaining
the full set of four-quark interactions shows explicitly how a running axial
coupling can move the phase boundary~\cite{Braun:2020mhk}.  Together these
results reinforce the point summarized in
Table~\ref{tab:transition-scenarios}: the value of an axial-breaking coupling
at one scale does not determine the transition.  Its flow together with the
other interactions decides whether the long-distance theory approaches a
stable fixed point or flows away from it, as expected for a first-order
transition.  This is why the relevant test must be made
near $T_c$, on length scales comparable to the growing correlation length and
in the light-quark limit.  Suppression observed only far above the transition
does not determine its order or universality class.

For $N_f\geq3$, the phase-transition question and the diagnostic question
should be separated.  Consider first the effective potential near the
transition.  The leading
determinant interaction is cubic for $N_f=3$ and quartic for $N_f=4$; it no
longer enters the quadratic mass matrix about $\Phi=0$, but it can still alter
the nonlinear renormalization-group flow and the order of the transition.
Moreover, the determinant in Eq.~(\ref{eq:lgw}) is only the lowest member of a
more general anomalous potential.  Schematically, one may write
\begin{equation}
 V_A(\Phi)=-\sum_{n\geq1}\sum_{j,k\geq0}\xi_n^{(j,k)}
 [\mathrm{Tr}(\Phi^\dagger\Phi)]^j
 [\mathrm{Tr}((\Phi^\dagger\Phi)^2)]^k
 \frac{(\det\Phi)^n+(\det\Phi^\dagger)^n}{2}+\cdots .
 \label{eq:anomalous-tower}
\end{equation}
The index $n$ labels the axial charge carried by the interaction; the omitted
terms contain other chirally invariant combinations.  Pisarski and Rennecke
emphasized that higher-order couplings can contribute to the large vacuum
$\eta'$ mass when $\langle\Phi\rangle\neq0$, whereas their contributions are
suppressed as the condensate vanishes at a chiral transition.  A heavy $\eta'$
at zero temperature therefore does not, by itself, determine the leading
anomalous coupling at $T_c$.  For three massless flavors, they argued that a
sufficiently small cubic coupling could substantially reduce the expected
first-order region, and conjectured that all anomalous couplings may vanish at
$T_c$ if the transition is continuous.  This provides a concrete target for
first-principles QCD calculations, but remains a conjecture rather than a
consequence of the anomaly
~\cite{Pisarski:2024esv}.

Continuum functional and effective calculations do not yet give a unique
many-flavor phase boundary.  Depending on the degrees of freedom, truncation,
and ultraviolet input, they find a possible three-flavor fixed point, weak or
first-order behavior, or a continuous transition in restricted regions of
coupling space~\cite{Mitter:2013fxa,Fejos:2022mso,Fejos:2024bgl,
Giacosa:2024orp,Bernhardt:2023hpr}.  Their common contribution is to identify
how a scale-dependent anomalous interaction can move or weaken the transition;
the alternatives remain targets for first-principles QCD calculations rather
than a settled phase diagram.

Constraints also follow from 't~Hooft anomaly matching: a long-distance theory
must reproduce the anomalies of QCD's exact global symmetries.  For massless
quarks at the Roberge--Weiss point of imaginary baryon chemical potential, a
mixed anomaly between non-singlet chiral symmetry and a discrete symmetry
associated with confinement restricts the phase structure
~\cite{Yonekura:2019vyz}.  Assuming a continuous line of chiral transitions
reached by tuning temperature as the imaginary chemical potential varies,
Chen et al. derive further restrictions on critical descriptions, including
at zero chemical potential~\cite{Chen:2026jla}.  These constraints do not
directly determine the strength of the $U_A(1)$-breaking interactions.

Direct tests of axial breaking complement these studies of the phase boundary.
Their form depends on the number of light flavors.  For two flavors, the
difference $\Delta_{\pi\delta}\equiv\chi_\pi-\chi_\delta$ between isovector
pseudoscalar and scalar susceptibilities provides a natural two-point
diagnostic.  For three or more massless flavors, two-point degeneracy in the
chirally symmetric phase can coexist with axial breaking in higher-point
correlations~\cite{Birse:1996dx}.  Section~\ref{sec:diagnostics} develops
these diagnostics and clarifies the relevant quark-mass limits.

\subsection{Topology, axial breaking and high-temperature limits}

Two questions are often conflated: how topology constrains mesonic $U_A(1)$
observables near the transition, and when topological fluctuations admit a
dilute semiclassical description.  These questions are related but not
interchangeable.  The first can be addressed without assuming a dilute
instanton gas.  Suppose that non-singlet chiral symmetry is restored and that
the free energy is analytic in a sufficiently small two-flavor mass matrix
$M$.  Through quadratic order one may write
\begin{equation}
 f(T,M,\theta)=f_0-f_2\,\mathrm{tr}(M^\dagger M)
 -f_A\left(e^{i\theta}\det M+e^{-i\theta}\det M^\dagger\right)
 +O(M^4).
 \label{eq:mass-expanded-free-energy}
\end{equation}
The coefficient $f_A$ measures the leading axial-breaking response in this
mass expansion.  We denote the topological susceptibility by $\chi_t$.  For
two flavors $f_A$ gives, in the conventions of
Refs.~\cite{Kanazawa:2014cua,Kanazawa:2015xna},
\begin{equation}
 \Delta_{\pi\delta}=8f_A+O(M^2),\qquad
 \chi_t=2f_A m_um_d+O(M^4).
 \label{eq:fA-observables}
\end{equation}
For degenerate light quarks, Eq.~(\ref{eq:fA-observables}) gives
$\Delta_{\pi\delta}=4\chi_t/m_l^2+O(m_l^2)$.  The same mass dependence follows,
without assuming the analytic free-energy expansion, from anomalous and
non-singlet chiral Ward identities once non-singlet chiral symmetry is
restored; the full relation, including $\chi_{5,\mathrm{disc}}$, is given in
Eq.~(\ref{eq:ward-topology})
~\cite{HotQCD:2012vvd,Aoki:2021qws}.  Thus $\chi_t\to0$ alone does not
establish effective restoration in this two-point sector: the required
condition is $\chi_t/m_l^2\to0$.

Under the same assumptions, the $\theta$-dependent part of
Eq.~(\ref{eq:mass-expanded-free-energy}) admits a Poisson representation in
terms of objects with charges $+1$ and $-1$, called quasi-instantons.  This is
a rewriting of the analytic free energy, not the dilute instanton gas
approximation: quantum effects are already contained in $f_A$, and no gas of
bare semiclassical instantons is assumed~\cite{Kanazawa:2014cua}.  The
representation relies on restored non-singlet symmetry and a regular mass
expansion, assumptions that can fail at a second-order chiral critical point.
Its Poisson counting also says nothing about the statistics of individual
Dirac eigenvalues, which may remain strongly correlated
~\cite{Ding:2020xlj}.

A semiclassical dilute-gas description requires additional conditions.
Thermal screening suppresses large instantons, and at sufficiently high
temperature the running coupling is small at the scale $T$.  An ensemble of
rare, weakly interacting instantons and anti-instantons should then provide a
useful description.  The dilute instanton gas approximation (DIGA) predicts a
rapid, approximately power-law decrease of the topological
susceptibility~\cite{Gross:1980br}.  For
$N_c=3$ and three light flavors, its leading temperature dependence is
schematically close to $\chi_t\propto T^{-8}$, multiplied by running-coupling,
quark-mass, and threshold corrections~\cite{Gross:1980br}.  The precise
exponent over a finite temperature interval is not universal.

This asymptotic reasoning should not be carried down automatically to
$T_{\rm pc}$.  Near the crossover the coupling is strong, topological objects
can interact, and DIGA is not controlled.  First-principles lattice QCD
calculations find a slower decrease of $\chi_t(T)$ near the crossover than
DIGA predicts, while its temperature dependence becomes increasingly
compatible with DIGA at higher temperatures
~\cite{Petreczky:2016vrs,Borsanyi:2016ksw}.  These results are reviewed in
Section~\ref{sec:topology-spectrum-evidence}.  DIGA therefore provides a useful asymptotic
endpoint, but not an automatic description of the transition region.

Independent high-temperature analyses based on correlation functions reach a
compatible asymptotic conclusion.  Debye screening of the Chern--Simons
current implies that selected axial-current correlators cannot behave as if an
exactly conserved $U_A(1)$ current had emerged~\cite{Laine:2003bd,Kanazawa:2015xna}, while explicit
instanton--anti-instanton chains generate a scalar--pseudoscalar screening
splitting that falls as a high power of $\Lambda_{\rm QCD}/T$
~\cite{Dunne:2010gd}.  These
arguments show how axial effects can be nonzero but parametrically small at
asymptotically high temperature.  They do not fix their magnitude in the
strongly coupled crossover region.

Although $\chi_t$ alone does not determine effective $U_A(1)$ restoration, it
has an important independent application in axion cosmology.  The axion acts
as a dynamical vacuum angle, so its QCD-induced mass is controlled by the
curvature of the QCD free energy with respect to $\theta$:
\begin{equation}
 \chi_t(T)=\frac{\langle Q^2\rangle}{V_4}
 =\left.\frac{\partial^2 f(\theta,T)}{\partial\theta^2}\right|_{\theta=0},
 \label{eq:top-susceptibility}
\end{equation}
This gives the temperature-dependent mass relation
$m_a^2(T)f_a^2=\chi_t(T)$, where $f_a$ is the axion decay constant~\cite{Bonanno:2026zzf}.
The cosmological abundance of axion dark matter is
therefore sensitive to QCD topology~\cite{Borsanyi:2016ksw,
Petreczky:2016vrs}.  This relation connects lattice QCD with particle physics
and cosmology.  The distinction between the two questions is important: axion
cosmology primarily needs the full $\theta$ dependence and $\chi_t(T)$ at
physical masses, whereas the chiral-transition problem asks whether
$U_A(1)$-violating infrared correlators survive as $m_l\to0$.

The conceptual separation is now clear.  Near a possible chiral critical
point, the question is whether axial breaking remains relevant on the longest
length scales.  Far above the transition, DIGA provides an asymptotic endpoint
but not a description of the crossover.  At physical masses, $\chi_t(T)$ is
essential for axion cosmology, whereas the chiral-limit symmetry question
depends on its mass scaling and on axial correlation functions.  A lattice
comparison must therefore specify the quark masses, volume, cutoff and fermion
formulation, observable, and order of limits.

\section{Historical development of the finite-temperature problem}
\label{sec:history}

The thermal problem developed through three changes of emphasis: from the
high-temperature suppression of instantons to critical behavior near the
chiral phase transition, from topological objects to their fermionic signatures, and from
individual symmetry tests to relations among observables.
Table~\ref{tab:history} records the main developments.

\begin{table}[htbp]
\centering
\begin{tabular}{p{0.12\textwidth}p{0.80\textwidth}}
\toprule
Period & Development \\
\midrule
1969--1980 & The axial anomaly~\cite{Adler:1969gk,Bell:1969ts,Adler:1969er,Bardeen:1969md}, the BPST instanton~\cite{Belavin:1975fg}, finite-temperature gauge solutions~\cite{Harrington:1978ve}, the 't Hooft vertex~\cite{tHooft:1976rip,tHooft:1976snw}, the path-integral derivation~\cite{Fujikawa:1979ay,Fujikawa:1980eg}, and the Witten--Veneziano and Banks--Casher relations~\cite{Witten:1979vv,Veneziano:1979ec,Banks:1979yr} connect current nonconservation, topology, hadron phenomenology, and the Dirac spectrum. \\
1981--1989 & High-temperature semiclassics predicts suppression of large instantons~\cite{Gross:1980br}; the anomaly coefficient remains temperature independent~\cite{Itoyama:1982up}; the Nielsen--Ninomiya theorem and Ginsparg--Wilson relation frame the lattice chiral problem~\cite{Nielsen:1981hk,Ginsparg:1981bj}; Pisarski and Wilczek connect the anomaly to the universality of the chiral phase transition~\cite{Pisarski:1983ms}.  Interacting-instanton models further connect topology to chiral symmetry breaking and thermal restoration~\cite{Shuryak:1987ja,Nowak:1989jd}. \\
1990--1999 & Finite-volume spectral sum rules relate topology,
quark masses, and the low Dirac spectrum~\cite{Leutwyler:1992yt}.
Instanton-molecule models~\cite{Shuryak:1993ee,Schafer:1994nv},
continuum constraints~\cite{Cohen:1996ng,Birse:1996dx},
and analyses of explicit topological
sectors~\cite{Lee:1996zy,Evans:1996wf} develop distinct pictures of
axial breaking above $T_c$. Domain-wall and overlap fermions supply lattice formulations with controlled chiral symmetry~\cite{Kaplan:1992bt,Shamir:1993zy,Furman:1994ky,Neuberger:1997fp,Luscher:1998pqa}, while early staggered studies
probe the low spectrum and
partner-channel observables~\cite{Chandrasekharan:1995gt,Bernard:1996iz,
	Kogut:1998rh,Chandrasekharan:1998yx}. \\
2000--2006 & Full-QCD calculations follow topology~\cite{Alles:2000cg} and axial-partner channels across the transition, including exploratory domain-wall simulations~\cite{Vranas:1999dg,Chen:2000zu}.  
Fixed-topology analyses~\cite{Brower:2003yx} and Ward
identities~\cite{Marchi:2003wq} refine the theoretical framework;
renormalization-group and effective-model analyses sharpen the
connection with criticality~\cite{Butti:2003nu,Calabrese:2004uk,
	Lenaghan:2000kr},
while screening calculations provide high-temperature
benchmarks~\cite{Laine:2003bd}.
Continuum finite-size scaling establishes a smooth chiral crossover
at physical quark masses~\cite{Aoki:2006we}. \\
2007--2012 & Fixed-topology expansions~\cite{Aoki:2007ka} and domain-wall thermodynamics improve control of topology and chiral symmetry~\cite{HotQCD:2012vvd}.  Improved-staggered screening correlators~\cite{Cheng:2010fe} and dynamical-HISQ spectra provide direct thermal tests~\cite{Ohno:2012br}, while eigenvector statistics reveal localized low modes and a mobility edge~\cite{Gavai:2008xe,Kovacs:2012zq}. \\
2013--2020 & Quasi-instanton and finite-temperature sum-rule analyses
relate topology to low-mode correlations
~\cite{Kanazawa:2014cua,Kanazawa:2015xna}.
Domain-wall, overlap, and reweighted calculations probe the
low-lying Dirac modes
~\cite{Chiu:2013wwa,Cossu:2013uua,Buchoff:2013nra,
	Cossu:2016scb,Tomiya:2016jwr,Suzuki:2020rla}.
Localization studies support three-dimensional unitary Anderson
criticality~\cite{Giordano:2013taa,Ujfalusi:2015nha}. Gradient flow and fermionic observables probe topological fluctuations
~\cite{Taniguchi:2016tjc,Burger:2018fvb}.\\
2021--present & Quark-mass derivatives of Dirac spectrum and connected eigenvalue correlations~\cite{Ding:2020xlj,Ding:2023oxy}, continuum low-mode densities~\cite{Alexandru:2024tel}, axial condensates~\cite{Carabba:2021xmc}, localization~\cite{Giordano:2024jnc}, and screening correlators~\cite{Aoki:2025mue} broaden the tests~\cite{Kaczmarek:2021ser,JLQCD:2024xey}; the analyticity assumptions~\cite{Azcoiti:2021gst,Azcoiti:2023xvu,Giordano:2025vbb,Aoki:2026hzl} and flavor dependence~\cite{Pisarski:2024esv} of earlier conclusions~\cite{Aoki:2012yj} are also being reassessed. \\
\bottomrule
\end{tabular}
\caption{Selected milestones in the study of the axial anomaly and hot QCD.
The entries mark conceptual turns rather than an exhaustive priority list.}
\label{tab:history}
\end{table}

\subsection{From high-temperature suppression to the chiral phase-transition question}

The early finite-temperature argument offered a simple asymptotic picture.
Thermal screening suppresses large instantons and asymptotic freedom weakens
the coupling at the scale $T$, so the density of semiclassical topological
objects falls rapidly at sufficiently high temperature~\cite{Gross:1980br}.
This established an important endpoint, but it did not determine the behavior
in the strongly coupled transition region.

Pisarski and Wilczek shifted attention from asymptotically high temperature to
the immediate vicinity of a possible critical point~\cite{Pisarski:1983ms}.
The question was no longer whether instantons are ultimately rare, but whether
the axial-breaking interaction remains relevant on the longest length scales
at $T_c$.  Its strength changes the symmetry and possible fixed-point structure
of the chiral effective theory. 

The 1990s literature developed two lines of reasoning.  In the
interacting-instanton picture, Shuryak stressed that restoration of non-singlet
chiral symmetry does not answer whether $U_A(1)$-sensitive channels become
degenerate at the same temperature~\cite{Shuryak:1993ee}.  The transition was
associated with a change from a disordered instanton liquid to correlated,
polarized instanton--anti-instanton molecules.  Such molecules can survive
above $T_c$, affect scalar and pseudoscalar propagation, and contribute little
to the net topological charge~\cite{Schafer:1994nv}.  Their effects on
fermion propagation must be distinguished from a surviving $U_A(1)$-breaking
interaction; neutral pairing alone does not establish the latter.
An instanton-liquid study focused directly on the axial anomaly found that its
effective strength could remain substantial near the chiral transition even
as restoration of non-singlet symmetry reorganized the $\eta$--$\eta'$ flavor
content~\cite{Schafer:1996hv}.

The second line used continuum inequalities and constraints on the infrared
Dirac spectrum.  Cohen argued, under regularity and analyticity assumptions,
that selected two-point functions in the chirally restored phase may exhibit
the larger $U(N_f)_L\times U(N_f)_R$ symmetry
~\cite{Cohen:1996ng,Cohen:1997hz}.  Lee and Hatsuda showed how topologically
nontrivial sectors qualify the strongest conclusion~\cite{Lee:1996zy}.
Evans, Hsu, and Schwetz, and Birse and collaborators, clarified the flavor
counting: for $N_f>2$, the anomaly may first appear in higher-point operators
without splitting selected thermal two-point functions
~\cite{Evans:1996wf,Birse:1996dx}.  These arguments established that the
observable, the flavor content, and the assumptions about the infrared
spectrum must all be stated.

In parallel with these theoretical developments, early staggered-fermion
calculations combined the Dirac spectrum with pion, scalar, and singlet
channels.  Chandrasekharan and Christ directly asked whether the hot spectrum
contained enough small eigenvalues to sustain axial breaking; subsequent
studies compared the $\pi$, $\delta$, $\sigma$, and $\eta'$ sectors and their
topology~\cite{Chandrasekharan:1995gt,
Kogut:1998rh,Chandrasekharan:1998yx}.  Bernard et al. found
non-singlet partners approaching degeneracy just above the crossover while
the $U_A(1)$-related splitting remained~\cite{Bernard:1996iz}.  Although
coarse lattices, taste breaking, heavier quarks, and uncontrolled chiral limits
restricted quantitative conclusions, these studies framed the first lattice
versions of the central questions.

This phase-transition thread has recently been revisited by Pisarski and Rennecke, who generalized 
the determinant interaction to the anomalous tower in Eq.~\eqref{eq:anomalous-tower} and examined 
how the full anomalous potential changes as the chiral order parameter decreases. Their proposed
tests for $N_f=1,\ldots,4$ offer new ways to address the original universality
question through first-principles QCD calculations~\cite{Pisarski:2024esv}.

\subsection{Spectral constraints and modern lattice strategies}

Alongside the phase-transition discussion, two developments enabled the
modern spectral program.  Leutwyler and Smilga
matched the finite-volume QCD partition function at fixed winding number to
its low-energy form and related inverse moments of the small Dirac eigenvalues
to the mass and topology dependence of the partition function
~\cite{Leutwyler:1992yt}. A separate line of work developed ways to
 represent quarks on a spacetime lattice while controlling violations of chiral symmetry.  The Nielsen--Ninomiya theorem and the
Ginsparg--Wilson relation defined the problem and its resolution~\cite{Nielsen:1981hk,Ginsparg:1981bj,Luscher:1998pqa}; domain-wall and
overlap fermions then made the lattice index theorem and the distinction
between exact and near-zero modes much cleaner~\cite{Kaplan:1992bt,Shamir:1993zy,Furman:1994ky,Neuberger:1997fp}.
An eigenmode analysis of M\"obius domain-wall fermions subsequently showed
that residual violations of the Ginsparg--Wilson relation can be enhanced in
the low-lying modes. Because the $U_A(1)$-sensitive susceptibility difference
$\chi_\pi-\chi_\delta$ is dominated by these modes, such violations can
substantially contaminate or even dominate its measured signal, particularly
on coarse lattices and at small quark masses~\cite{Cossu:2015kfa}.

Several theoretical analyses classified infrared spectra compatible with
restored non-singlet chiral symmetry.  Aoki, Fukaya, and Taniguchi derived
strong suppression within a regular, Taylor-expandable spectral framework
with $m_l^2$-analyticity and a specified order of limits
~\cite{Aoki:2012yj}.  Kanazawa and Yamamoto separated these assumptions and
related the chirally restored free energy to quasi-instantons, fixed-topology
sum rules, and two-level eigenvalue correlations
~\cite{Kanazawa:2014cua,Kanazawa:2015xna}.  Azcoiti examined nonanalytic mass
dependence, while recent work by Giordano and the ensuing comment and reply
sharpened singular alternatives and the scope of the analyticity assumptions
~\cite{Azcoiti:2021gst,Azcoiti:2023xvu,Giordano:2024jnc,
Giordano:2025shr,Giordano:2025fcr,Giordano:2025vbb,Aoki:2026hzl}.
These studies highlight how assumptions about the near-zero Dirac spectrum and 
the order of the infinite-volume and chiral limits shape conclusions about axial breaking.

The studies also broadened to include additional quark flavors and correlations among Dirac eigenvalues.  Higher-point
axial condensates and strange-sector Ward identities exposed the flavor
dependence~\cite{Carabba:2021xmc,GomezNicola:2020qxo}. 
Ding et al. related derivatives of the Dirac spectral density with respect to the 
light sea-quark mass to connected correlations among Dirac eigenvalues~\cite{Ding:2020xlj}.  The generalized Banks--Casher relation subsequently
connected cumulants of the chiral order parameter to spectral correlations at
the origin~\cite{Ding:2023oxy}.

Three broad lattice strategies emerged.  Domain-wall, overlap, and
overlap-reweighted studies tested axial observables with increasingly direct
control of chiral symmetry, including work by the HotQCD and JLQCD
collaborations and studies using optimal domain-wall fermions
~\cite{HotQCD:2012vvd,Buchoff:2013nra,Tomiya:2016jwr,Suzuki:2020rla,
Aoki:2020noz,Aoki:2022ebi,JLQCD:2024xey,Chen:2022fid,Chiu:2023hnm,
Chiu:2024jyz,Chiu:2024bqx}.  Improved-staggered and mixed-action studies
developed high-statistics tests of the infrared density, its mass dependence,
and its continuum behavior
~\cite{Ohno:2012br,Dick:2015twa,Kaczmarek:2021ser,Ding:2020xlj,
Kaczmarek:2023bxb,Alexandru:2024tel}, while Wilson-fermion calculations
provided correlator-based tests without directly determining the near-zero
density~\cite{Brandt:2016daq}, with recent results from the FASTSUM
collaboration~\cite{Aarts:2026kpq}.  Because these strategies differ
in sea and valence actions, quark masses, volumes, lattice spacings, and the
order of limits, their results must be compared through common observables and
controlled limits rather than formulation names alone.
Section~\ref{sec:diagnostics} introduces these common observables and explains the relations between them.

\section{How axial breaking is diagnosed}
\label{sec:diagnostics}

The historical developments above converged on a practical lesson: a statement
about effective $U_A(1)$ restoration is meaningful only after the observable
and the order of limits have been specified.  For the two-flavor problem,
meson susceptibilities, the Dirac spectrum, and the topological susceptibility
are connected by Ward identities, but they emphasize partner-channel
degeneracy, infrared eigenmodes, and net topological-charge fluctuations,
respectively.  We begin with the partner channels and the average Dirac
spectrum, then examine the additional information in quark-mass derivatives
and eigenvalue correlations, and finally turn to topology and further
correlation-function tests.

\subsection{Meson observables and the infrared Dirac spectrum}

We follow the operator and susceptibility normalization of
Ref.~\cite{Ding:2020xlj}.  Let $\psi=(u,d)^T$ with degenerate light-quark
mass $m_l$, and let $\tau^a$ be a Pauli matrix in flavor space.  Four scalar
and pseudoscalar bilinears are particularly useful:
\begin{align}
 \pi^a(x)&=\bar\psi(x)i\gamma_5\tau^a\psi(x), &
 \delta^a(x)&=\bar\psi(x)\tau^a\psi(x), \nonumber\\
 \sigma(x)&=\bar\psi(x)\psi(x), &
 \eta(x)&=\bar\psi(x)i\gamma_5\psi(x).
 \label{eq:meson-operators}
\end{align}
The $\pi$ and $\delta$ (often called $a_0$) are flavor non-singlets, whereas the $\sigma$ ($f_0$)
and $\eta$ are isosinglets whose correlators contain quark-line-disconnected
contractions.  These names label operator channels rather than individual
vacuum resonances.
Overall normalization conventions differ across the literature, but the
degeneracy relations do not.

\begin{figure}[htbp]
\centering
\begin{minipage}[c]{0.53\textwidth}
\centering
\definecolor{partnerBlue}{RGB}{38,78,130}
\definecolor{partnerPurple}{RGB}{137,62,113}
\definecolor{partnerOutline}{RGB}{149,161,175}
\definecolor{partnerFill}{RGB}{247,249,252}
\begin{tikzpicture}[
 channel/.style={draw=partnerOutline,fill=partnerFill,
                 rounded corners=2pt,line width=0.55pt,
                 minimum width=2.5cm,minimum height=1.02cm,
                 align=center,inner sep=4pt},
 chiral/.style={{Latex[length=1.65mm,width=1.1mm]}-{Latex[length=1.65mm,width=1.1mm]},
                draw=partnerBlue,line width=0.8pt,
                shorten <=1pt,shorten >=1pt},
 axial/.style={{Latex[length=1.65mm,width=1.1mm]}-{Latex[length=1.65mm,width=1.1mm]},
               draw=partnerPurple,line width=0.8pt,
               dash pattern=on 2.2pt off 1.7pt,
               shorten <=1pt,shorten >=1pt}]
 \node[channel] (pi) at (0,2.15)
   {{\large $\pi^a$}\\[1.5pt]
    {\color{black!85}\small $\chi_\pi$}};
 \node[channel] (sigma) at (4.05,2.15)
   {{\large $\sigma$}\\[1.5pt]
    {\color{black!85}\small $\chi_\delta+\chi_{\rm disc}$}};
 \node[channel] (delta) at (0,0)
   {{\large $\delta^a$}\\[1.5pt]
    {\color{black!85}\small $\chi_\delta$}};
 \node[channel] (eta) at (4.05,0)
   {{\large $\eta$}\\[1.5pt]
    {\color{black!85}\small $\chi_\pi-\chi_{5,\rm disc}$}};
 \draw[chiral] (pi.east) -- (sigma.west);
 \draw[chiral] (delta.east) -- (eta.west);
 \draw[axial] (pi.south) --
   node[left=3pt,text=partnerPurple,font=\footnotesize] {$U_A(1)$}
   (delta.north);
 \draw[axial] (sigma.south) --
   node[right=3pt,text=partnerPurple,font=\footnotesize] {$U_A(1)$}
   (eta.north);
 \node[text=partnerBlue,font=\footnotesize] at (2.025,2.95)
   {$SU(2)_L\times SU(2)_R$};
 \node[anchor=north west,font=\small] at (-1.38,3.14) {(a)};
\end{tikzpicture}
\end{minipage}\hfill
\begin{minipage}[c]{0.43\textwidth}
\centering
\begin{tikzpicture}[x=0.88cm,y=2.35cm]
 \draw[-{Latex}] (0,0) -- (5.15,0) node[right,font=\scriptsize] {$\lambda$};
 \draw[-{Latex}] (0,0) -- (0,1.18);
 \draw (-0.07,1) -- (0.07,1);
 \node[anchor=east,font=\scriptsize] at (-0.10,1) {$1$};
 \node[anchor=west,font=\scriptsize] at (0.12,1.25)
   {$K(\lambda,m_l)/K(0,m_l)$};
 \draw[blue,thick,samples=100,domain=0:4.9]
   plot (\x,{1/(1+(\x/1.35)^2)^2});
 \draw[magenta,thick,samples=100,domain=0:4.9]
   plot (\x,{1/(1+(\x/0.65)^2)^2});
 \draw[dashed] (0.65,0) -- (0.65,0.25);
 \draw[dashed] (1.35,0) -- (1.35,0.25);
 \node[magenta,anchor=north,font=\scriptsize] at (0.65,-0.025) {$m_1$};
 \node[blue,anchor=north,font=\scriptsize] at (1.35,-0.025) {$m_2$};
 \node[magenta,anchor=west,font=\scriptsize] at (1.55,0.86)
   {smaller $m_l=m_1$};
 \node[blue,anchor=west,font=\scriptsize] at (2.05,0.53)
   {larger $m_l=m_2$};
 \node[anchor=north west,font=\small] at (-0.42,1.55) {(b)};
\end{tikzpicture}
\end{minipage}
\caption{Two views of the two-flavor diagnostics.  (a) Solid
horizontal arrows connect non-singlet chiral partners; dashed vertical arrows
connect $U_A(1)$ partners.  The arrows identify transformation properties,
not an assumed degeneracy.  (b) The kernel
\mbox{$K(\lambda,m_l)=8m_l^2/(\lambda^2+m_l^2)^2$} in
Eq.~\eqref{eq:spectral-delta}, normalized to unit height for illustrative
masses $m_1<m_2$.  Dashed lines mark $\lambda=m_1,m_2$, where the normalized
kernel is $1/4$.  Lowering $m_l$ narrows the spectral window and raises the
unnormalized height, $K(0,m_l)=8/m_l^2$, so a small near-zero population can
leave a finite $\Delta_{\pi\delta}$.}
\label{fig:meson-partners-and-kernel}
\end{figure}
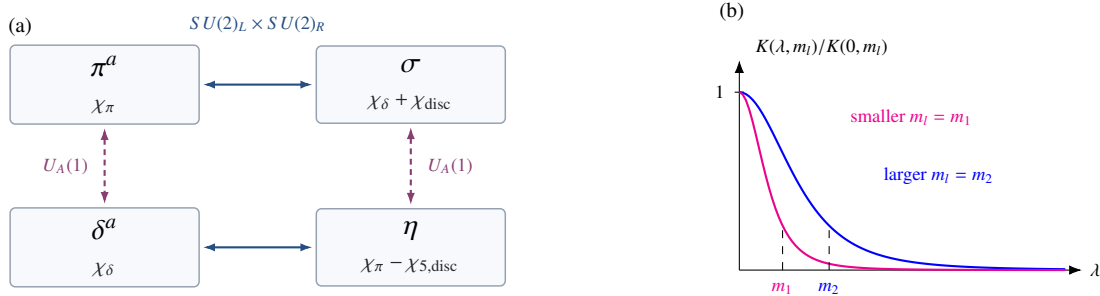

The four channels form two different sets of symmetry partners.  In terms of
their susceptibilities,
\begin{align}
 SU(2)_L\times SU(2)_R:&\qquad
 \chi_\pi=\chi_\sigma,\qquad \chi_\delta=\chi_\eta,
 \nonumber\\
 U_A(1):&\qquad
 \chi_\pi=\chi_\delta,\qquad \chi_\sigma=\chi_\eta.
 \label{eq:partner-pattern}
\end{align}
The first line follows when non-singlet chiral symmetry is restored in the
chiral limit.  The second line is the corresponding two-point criterion for
effective axial restoration.  Thus restoration of
$SU(2)_L\times SU(2)_R$ alone does not require the pion and $\delta$ channels
to agree.  If both lines hold, all four susceptibilities are degenerate.
Figure~\ref{fig:meson-partners-and-kernel}(a) summarizes these partner
relations, following Ref.~\cite{HotQCD:2012vvd}.

For an operator $X$, its thermal susceptibility is the spacetime integral
\begin{equation}
 \chi_X=\int_0^{1/T}d\tau\int d^3x\,
 \langle X(\tau,\mathbf{x})X(0)\rangle_{\rm conn/full},
 \label{eq:susceptibility-definition}
\end{equation}
Nonsinglet susceptibilities refer to one fixed isospin component $a$.
Here ``conn'' refers to quark-line-connected contractions in the nonsinglet
channels; ``full'' also includes disconnected contractions in the singlet
channels.  In the scalar isosinglet channel, the squared condensate
$\langle\sigma\rangle_T^2$ is subtracted from the correlator, while
quark-line-disconnected fluctuations are retained.  A widely used two-point
measure of axial breaking is
\begin{equation}
 \Delta_{\pi\delta}\equiv\chi_\pi-\chi_\delta.
 \label{eq:delta-pidelta}
\end{equation}
It is numerically attractive because both channels are flavor nonsinglets, 
so its calculation avoids quark-line-disconnected contractions, which are typically noisy in lattice calculations.
At nonzero $m_l$, explicit mass breaking and anomaly effects are mixed; the
relevant question is whether Eq.~(\ref{eq:delta-pidelta}) approaches zero
after the thermodynamic and chiral limits at a specified temperature.

The same four channels enter two useful integrated Ward identities.  Define
the positive magnitude of the light condensate by
$\Sigma_l\equiv-\langle\bar u u+\bar d d\rangle$, with the negative
condensate convention of Eq.~\eqref{eq:banks-casher}.  The
non-singlet identity gives
\begin{equation}
 \Sigma_l=m_l\chi_\pi.
 \label{eq:condensate-pion-ward}
\end{equation}
The integrated anomalous Ward identity relates the disconnected pseudoscalar
susceptibility to topology:
\begin{equation}
 \chi_{5,\mathrm{disc}}\equiv\chi_\pi-\chi_\eta,
 \qquad
 \chi_t=\frac{m_l^2}{4}\chi_{5,\mathrm{disc}}.
 \label{eq:topology-pseudoscalar-ward}
\end{equation}
The factor $1/4$ follows from the two-flavor operator normalization in
Eq.~\eqref{eq:meson-operators}.  We also define the disconnected chiral
susceptibility, $\chi_{\rm disc}\equiv\chi_\sigma-\chi_\delta$, which measures
fluctuations of the light-quark condensate across gauge configurations.  The restored
non-singlet relations in Eq.~\eqref{eq:partner-pattern} imply
\begin{equation}
 \Delta_{\pi\delta}=\chi_{\rm disc}
 =\chi_{5,\mathrm{disc}}=\frac{4\chi_t}{m_l^2}.
 \label{eq:ward-topology}
\end{equation}
Only the last step uses the anomalous Ward identity; the equalities to
$\Delta_{\pi\delta}$ additionally require restored non-singlet chiral
symmetry~\cite{HotQCD:2012vvd,Aoki:2021qws}.

The massless Euclidean Dirac operator has paired eigenvalues $\pm i\lambda$.
Let $\rho(\lambda;m_l,T)$ be the ensemble-averaged density of positive
eigenvalues of a single-flavor Dirac operator per unit four-volume; the
two light flavors are included in the ensemble weight, not counted again
in $\rho$.  After the thermodynamic limit, the above bilinear conventions give
the spectral representation
\begin{equation}
 \Delta_{\pi\delta}
 =8m_l^2\int_0^\infty d\lambda\,
 \frac{\rho(\lambda;m_l,T)}{(\lambda^2+m_l^2)^2}.
 \label{eq:spectral-delta}
\end{equation}
The factor of eight follows from the two-flavor normalization and the
restriction to positive eigenvalues~\cite{Ding:2020xlj,Kanazawa:2015xna}.
Contributions from exact zero modes at finite volume are discussed in
Section~\ref{sec:order-of-limits}.

Equation~(\ref{eq:spectral-delta}) turns the light-quark mass into a spectral
resolution scale.  The kernel is concentrated at $\lambda\lesssim m_l$; as
$m_l$ decreases, the observable examines an ever narrower region around the
origin.  Modes far above $m_l$ are strongly suppressed.  Panel (b) of
Fig.~\ref{fig:meson-partners-and-kernel} illustrates this narrowing spectral
window by displaying the kernel normalized to unit height.

\FloatBarrier

The spectral density specifies the distribution of Dirac eigenvalues, but
does not describe the spatial structure of their eigenfunctions.  A localized
mode remains concentrated within a finite spatial region as the volume
increases, whereas an extended mode spreads through the system.  Finite-size
scaling tests this distinction by tracking the effective volume occupied by
each mode.  A mobility edge separates localized and extended regions of the
spectrum.  Neighboring eigenvalue spacings, rescaled by their local mean,
typically show uncorrelated (Poisson) statistics for localized modes and
random-matrix level repulsion for extended modes
~\cite{Kovacs:2012zq,Giordano:2013taa}.  Near the mobility edge, multifractal
analysis examines how moments of the mode density scale with system size
~\cite{Ujfalusi:2015nha}.  Localization characterizes this spatial structure, but
does not by itself establish axial restoration.  The corresponding
measurements are reviewed in Section~\ref{sec:screening-evidence}.

To illustrate how the near-zero spectral density controls the
chiral limit, assume a leading behavior $\rho(\lambda)\simeq c\lambda^\alpha$
with negligible quark-mass dependence in that window.  For $0\leq\alpha<3$,
the infrared contribution scales as\footnote{For $\alpha=3$, the infrared
contribution behaves as $m_l^2\ln(\Lambda/m_l)$ and also vanishes as
$m_l\to0$; $\Lambda$ is a fixed upper cutoff on the infrared region.}
\begin{equation}
 \Delta_{\pi\delta}^{\rm IR}\sim m_l^{\alpha-1}.
 \label{eq:alpha-scaling}
\end{equation}
Within this approximation, a linear density ($\alpha=1$) leaves a finite
infrared contribution even though $\rho(0)=0$.  For $1<\alpha<3$, the stronger
suppression near the origin makes this contribution vanish in the chiral
limit.

This simple power-law description can fail when a near-zero peak narrows as
the quark mass decreases.  The density $\rho(\lambda;m_l)$ may then vanish at
every fixed nonzero $\lambda$ in the chiral limit while its contribution to
Eq.~\eqref{eq:spectral-delta} remains finite.  A schematic example is supplied
by instanton-induced near-zero modes.  For $N_f$ degenerate light flavors, the
fermion determinant gives their density an $m_l^{N_f}$ suppression, leading to
$\rho_{\rm nz}(\lambda;m_l)\simeq c(T)m_l^{N_f}\delta(\lambda)$.  The delta
function idealizes a narrow peak whose interaction-induced width is neglected.
For $N_f=2$, its $m_l^2$ weight is compensated by the infrared kernel, so a
nonzero chiral-limit value of $\Delta_{\pi\delta}$ can remain
~\cite{Gross:1980br,HotQCD:2012vvd}.  The more general low-energy form
$c_0m_l^2\delta(\lambda)+c_1|\lambda|+c_2m_l+\cdots$ shows how several terms
can contribute to $\Delta_{\pi\delta}$~\cite{Buchoff:2013nra}.  Thus neither
pointwise vanishing away from the origin nor a shrinking integrated weight
alone establishes effective restoration.  Spectra at several masses are more
informative than the presence or absence of a gap on one ensemble.

These examples divide the possible chiral-limit behavior into regular and
singular classes.  In the regular class, $\rho(\lambda;m_l)$ has a Taylor
expansion in $\lambda$ with a radius that remains nonzero as $m_l\to0$,
mass-independent gauge-field observables are analytic in $m_l^2$, and the
thermodynamic limit precedes the chiral limit.  Together with restored
non-singlet chiral symmetry, these assumptions strongly suppress the leading
near-zero terms and remove axial breaking from a class of integrated
correlators~\cite{Aoki:2012yj,Kanazawa:2015xna,Giordano:2025vbb,Aoki:2026hzl}.  This is a conditional result
for this regular spectral class; a narrowing distribution such as
$m_l^2\delta(\lambda)$ lies outside it.

Within these analyses, surviving two-point axial breaking requires regularity
to fail in a specified way.  One possibility is a narrow spectral peak,
represented schematically by $m_l^2\delta(\lambda)$, compatible with an
analytic quark-mass dependence of the free energy and nontrivial $\theta$
dependence.
Under finite scalar and pseudoscalar susceptibilities, this behavior is tied
to a singular connected two-eigenvalue correlation and delocalized near-zero
modes~\cite{Giordano:2024jnc,Giordano:2025shr,Giordano:2025fcr,
Kovacs:2023vzi}.  A second possibility is nonanalytic quark-mass dependence
of the free energy.  In the scenario discussed by Azcoiti, surviving axial
breaking in the chirally restored phase is accompanied by divergent scalar
and pseudoscalar correlation lengths as $m_l\to0$, corresponding to modes
whose masses vanish in that limit.  The regular free-energy expansion in
Eq.~\eqref{eq:mass-expanded-free-energy} cannot describe this nonanalytic
mass response.  The two-flavor Schwinger model, where
$\chi_t\propto m^{4/3}$, illustrates this possibility in lower dimensions but
does not determine four-dimensional QCD
~\cite{Azcoiti:2021gst,Azcoiti:2023xvu}.  These are allowed limiting scenarios,
not conclusions about which one QCD realizes.

The four-channel map can also be reorganized mode by mode.  On two-flavor
ensembles reweighted from a M\"obius domain-wall to an overlap determinant,
the axial-sensitive pieces accounted for most of the subtracted connected and
disconnected susceptibilities under the conditions studied
~\cite{Aoki:2021qws}.  This is a decomposition of the same partner relations,
not an additional restoration criterion.  More generally, the average density
at one quark mass cannot distinguish a regular spectrum from a narrowing,
correlated contribution.  The required information lies in quark-mass
derivatives of $\rho$ and in connected correlations among eigenvalues.

\subsection{Mass derivatives, eigenvalue correlations, and the chiral limit}
\label{subsec:spectral-correlations}

Equation~\eqref{eq:spectral-delta} uses the ensemble-averaged density at each
quark mass.  To determine how a near-zero feature changes with the sea-quark
mass, one differentiates this average.  Because the fermion determinant sets
the statistical weight of each gauge field, the derivative generates
connected correlations among its Dirac eigenvalues.  For a configuration
$U$, define the microscopic density of positive modes by
\begin{equation}
 \rho_U(\lambda)=\sum_j\delta(\lambda-\lambda_j[U]),
 \qquad
 \delta\rho_U(\lambda)=\rho_U(\lambda)-\langle\rho_U(\lambda)\rangle,
\end{equation}
and its connected two-point correlation by
$C_2(\lambda,\lambda')=\langle\delta\rho_U(\lambda)
\delta\rho_U(\lambda')\rangle$.  A finite-temperature Banks--Casher-type
relation makes
the role of two-level correlations particularly explicit.  If $T_2$ denotes
the deviation of the connected two-level correlator from its Poisson form
and varies only on spectral scales large compared with $m_l$, 
then the analytic high-temperature mass expansion gives, for two flavors and
in the normalization of Ref.~\cite{Kanazawa:2015xna},
\begin{equation}
 T_2(0,0)=\frac{2}{\pi^2}f_A.
 \label{eq:kanazawa-banks-casher}
\end{equation}
The ordinary Banks--Casher relation involves the one-level density.  Equation
(\ref{eq:kanazawa-banks-casher}) instead places the leading axial coupling in
a connected two-level quantity, showing why the average density alone need
not contain all anomaly information.  Its precise prefactor depends on the
spectral and flavor normalization.

Differentiating the fermion determinant with respect to the light sea-quark
mass gives, for two degenerate light flavors~\cite{Ding:2020xlj},
\begin{equation}
 V_4\frac{\partial\rho(\lambda;m_l)}{\partial m_l}
 =4m_l\int_0^\infty d\lambda'\,
 \frac{C_2(\lambda,\lambda';m_l)}{\lambda'^2+m_l^2}.
 \label{eq:rho-mass-derivative}
\end{equation}
Here $V_4=V/T$ and $\rho=\langle\rho_U\rangle/V_4$; factors change if the
density is normalized differently.  
Repeated differentiation relates
$\partial^n\rho/\partial m_l^n$ to connected correlations involving up to
$n+1$ eigenvalues.  
These identities allow the quark-mass derivatives of $\rho$ to be
computed directly from connected eigenvalue correlations measured
over an ensemble of gauge configurations generated at the same
light-quark mass $m_l$, without requiring additional ensembles at
nearby quark masses~\cite{Ding:2020xlj}.
Computing these derivatives at nonzero quark mass does not require regularity
or finite-susceptibility assumptions in the chiral limit; their mass dependence
provides a way to test those assumptions.

Mass derivatives also help disentangle contributions with different quark-mass
dependences.  For the schematic low-energy form
$\rho\simeq c_0m_l^2\delta(\lambda)+c_1|\lambda|+c_2m_l+\cdots$
introduced above, take the $c_i$ to be mass independent.
Then $\partial\rho/\partial m_l\simeq2c_0m_l\delta(\lambda)+c_2+\cdots$,
whereas $\partial^2\rho/\partial m_l^2\simeq2c_0\delta(\lambda)+\cdots$.
The first derivative removes the mass-independent $c_1|\lambda|$ term; the
second also removes $c_2m_l$, isolating the quadratic peak contribution among
these terms.  Comparing the density with its mass derivatives therefore provides a more direct
way to disentangle contributions with different powers of $m_l$, which are
superposed in $\rho(\lambda;m_l)$ and can become increasingly difficult to
resolve directly as mass-suppressed terms vanish toward the chiral limit, even
though they may still produce a finite chiral-limit
$\Delta_{\pi\delta}$~\cite{Buchoff:2013nra,Ding:2020xlj}.

A generalized Banks--Casher relation connects the average spectral density
and its connected correlations to the cumulants of the chiral condensate.  Let
$\kappa_n[X]=(T/V)\langle X^n\rangle_c$ denote the volume-normalized $n$th
connected cumulant of an extensive variable $X$.  With the two-flavor and
positive-eigenvalue conventions of Ref.~\cite{Ding:2023oxy}, the chiral limit
gives
\begin{equation}
 \lim_{m_l\to0}\kappa_n[\bar\psi\psi]
   =(2\pi)^n\kappa_n[\rho_U(0)],\qquad n\geq1,
 \label{eq:generalized-banks-casher}
\end{equation}
where $\bar\psi\psi$ denotes the spacetime-integrated two-flavor condensate.
For $n=1$, Eq.~(\ref{eq:generalized-banks-casher}) reduces to the usual
Banks--Casher relation after matching the sign and flavor normalizations.  For
$n\geq2$, it states that fluctuations of the chiral order parameter are
encoded in connected correlations among $n$ eigenvalues at the origin.  At
nonzero $m_l$, the delta-function support at $\lambda=0$ is broadened by the
kernel $4m_l/(\lambda^2+m_l^2)$, and the condensate cumulants become integrals
over the corresponding connected spectral correlations.

Near a continuous chiral phase transition, Eq.~\eqref{eq:generalized-banks-casher}
connects the infrared spectrum to the universal fluctuations discussed in
Section~\ref{sec:chiral-transition}.  Let $t$ be the reduced temperature,
$h\propto m_l$ the symmetry-breaking field, and
$z=t/h^{1/(\beta\delta)}$.  Here $\beta$ and $\delta$ are universal
order-parameter critical exponents of the relevant universality class:
$\Sigma_l\propto(-t)^\beta$ as $t\to0^-$ at $h=0$,
while $\Sigma_l\propto h^{1/\delta}$ at $t=0$.
The singular contribution to the $n$th cumulant
then scales schematically as
\begin{equation}
 \kappa_n(t,h)=h^{1/\delta-n+1}f_n(z).
 \label{eq:cumulant-scaling}
\end{equation}
The function $f_n(z)$ is the universal scaling function for the $n$th
condensate cumulant; $f_1$ describes the order parameter and $f_2$ its
susceptibility.
At the nonzero lattice spacing of the staggered spectral calculation, the
correlations entering Eq.~\eqref{eq:generalized-banks-casher} follow the
expected $O(2)$ scaling and encode the same macroscopic critical behavior
~\cite{Ding:2023oxy}.  This microscopic connection does not by itself settle
the axial question: chiral cumulants and $\Delta_{\pi\delta}$ are different
weighted combinations of the infrared spectral correlations.

The same mass derivative enters the disconnected chiral susceptibility; see
Refs.~\cite{Smilga:1995qf,HotQCD:2012vvd,Ding:2020xlj}.  With the conventions used
here,
\begin{equation}
 \chi_{\rm disc}=4m_l\int_0^\infty d\lambda\,
 \frac{\partial\rho(\lambda;m_l)/\partial m_l}
      {\lambda^2+m_l^2}.
 \label{eq:chidisc-rho-derivative}
\end{equation}
When non-singlet chiral symmetry is restored, the partner relations in
Eq.~\eqref{eq:partner-pattern} give
$\chi_{\rm disc}=\Delta_{\pi\delta}$.  Equations
~\eqref{eq:spectral-delta} and \eqref{eq:chidisc-rho-derivative} must then
agree, providing a direct check that the amount of infrared weight and its
mass response are mutually consistent.  
Both $\rho$ and its mass derivative can be evaluated on the same
ensemble of gauge configurations generated at one sea-quark mass,
although several masses and volumes are still needed to control
the chiral and thermodynamic limits.

\subsection{Topology and correlation-function tests}
\label{subsec:diagnostics_topology}

Topology provides a global diagnostic because \(q(x)\) is the anomalous term in 
Eq.~\eqref{eq:anomalousward} and its spacetime integral is the charge \(Q\) in Eq.~\eqref{eq:topological-charge}. 
The charge variance defines \(\chi_t\) through Eq.~\eqref{eq:top-susceptibility}. In two-flavor QCD, the anomalous Ward identity directly relates \(\chi_t\) 
to the disconnected pseudoscalar susceptibility \(\chi_{5,\mathrm{disc}}\), as in Eq.~\eqref{eq:topology-pseudoscalar-ward}. 
Its further identification with \(\Delta_{\pi\delta}\) and \(\chi_{\mathrm{disc}}\) requires the restored non-singlet chiral-symmetry relations entering Eq.~\eqref{eq:ward-topology}. 
We first outline the definitions and interpretation of topological observables, then examine what meson correlation functions reveal about propagation.

The gluonic determination evaluates Eq.~\eqref{eq:topological-charge} after
ultraviolet fluctuations have been smoothed, most commonly by gradient flow~\cite{Luscher:2010iy}.
A fermionic definition follows from the index theorem in
Eq.~\eqref{eq:index-theorem}: with a chirally
symmetric Dirac operator one counts its exact zero modes,
\begin{equation}
 Q_{\rm f}=n_L-n_R,\qquad
 \chi_t^{\rm f}=\frac{\langle Q_{\rm f}^2\rangle}{V_4}.
 \label{eq:fermionic-top-susceptibility}
\end{equation}
The overall sign convention for $Q_{\rm f}$ does not affect $\chi_t^{\rm f}$.
For a Ginsparg--Wilson operator, such as the overlap operator, $Q_{\rm f}$ is integer-valued at
nonzero lattice spacing~\cite{Neuberger:1997fp,Luscher:1998pqa}.  The gluonic
and fermionic definitions need not agree configuration by configuration at
finite lattice spacing, but they must give the same continuum susceptibility;
their comparison is therefore an important cross-check
~\cite{Luscher:2010iy,Alexandrou:2017hqw}.

A third determination uses the disconnected fluctuation of the singlet
pseudoscalar density.  Equation~\eqref{eq:topology-pseudoscalar-ward} gives
$\chi_t=m_l^2\chi_{5,\mathrm{disc}}/4$, while
Eq.~\eqref{eq:ward-topology} connects it to $\Delta_{\pi\delta}$ after
non-singlet chiral restoration.  This Ward-identity route is distinct from
counting zero modes, although both must agree with the gluonic definition in
the continuum.  Comparing all three determinations on the same ensembles can
expose residual chiral-symmetry or topology-measurement errors
~\cite{HotQCD:2012vvd,Aoki:2021qws}.

Higher cumulants of $Q$ characterize the shape of the topological-charge
distribution beyond its variance.  The coefficient $b_2$, determined by the fourth connected
moment of $Q$, characterizes the leading departure from quadratic
$\theta$ dependence.\footnote{At $\theta=0$,
$b_2=-(\langle Q^4\rangle-3\langle Q^2\rangle^2)/(12\langle Q^2\rangle)$.}
Its DIGA value, $b_2=-1/12$, provides a benchmark for the topological-charge
distribution, but agreement with this value alone does not establish
effective $U_A(1)$ restoration.  Numerical determinations are reviewed in
Section~\ref{sec:topology-spectrum-evidence}.

A limitation of $\chi_t$ is equally important: it measures fluctuations of
the net charge, not the number of topological objects paired into a $Q=0$
configuration\cite{Kanazawa:2015xna,Schafer:1994nv}.  Such pairs can affect low modes and meson correlators without
a large net-charge variance.  Whether an axial-breaking contribution survives
must be determined from those correlators and their Ward identities.  At physical
quark masses $\chi_t$ remains nonzero above the crossover
~\cite{Borsanyi:2016ksw,Petreczky:2016vrs,Chen:2022fid}, including at
temperatures where the $U_A(1)$-sensitive scalar--pseudoscalar screening-mass
splitting is numerically very small~\cite{Bazavov:2019www}.
This comparison motivates a closer look at the meson correlation functions
from which susceptibilities and screening masses are obtained.

The susceptibility in Eq.~\eqref{eq:susceptibility-definition} integrates a
two-point function over all Euclidean separations.  Keeping one spatial
separation, $z$, unintegrated instead defines the screening correlator
\begin{equation}
 G_X(z)=\int_0^{1/T}d\tau\int dx\,dy\,
 \langle X(\tau,x,y,z)X(0)\rangle.
 \label{eq:screening-correlator}
\end{equation}
With the same connected/full convention, Eq.~\eqref{eq:susceptibility-definition}
is recovered from Eq.~\eqref{eq:screening-correlator} through
$\chi_X=\int dz\,G_X(z)$.  At large $z$, $G_X(z)$ decays with a screening mass
$M_X$.  Degenerate $\pi$ and $\delta$ screening masses therefore test axial
breaking in the longest spatial correlation length.  Agreement of the full
correlator shapes and amplitudes is stronger evidence than agreement of fitted
masses alone.  Screening observables are physically intuitive and do not
require explicit eigenvalue calculations, but singlet partners involve noisy
disconnected diagrams and finite spatial extents can obscure the asymptotic
regime.

A different projection keeps the Euclidean-time separation and gives access to
thermal spectral functions.  The temporal extent is only $1/T$, 
so the physical time interval available to constrain the correlator becomes shorter as the temperature increases.
 Inferring real-time peaks or widths is an ill-posed inverse problem.  For the axial
question, direct comparison of symmetry-related Euclidean correlators is
usually more robust than attempting a detailed spectral reconstruction.

These observables emphasize different parts of the same two-point function.
The susceptibility weights all separations, the screening mass isolates the
slowest exponential at large spatial separation, and the temporal spectral
function describes the excitation structure.  Consequently, equality of two
partner susceptibilities and degeneracy of their screening masses need not
occur at the same temperature.  At nonzero quark mass, modes that contribute
substantially to an integrated susceptibility need not determine the longest
screening length, since correlators at separated points also depend on the
spatial structure of the eigenfunctions~\cite{Dick:2015twa}.

Comparisons of meson partners also extend to channels containing a strange
quark.  Chiral Ward
identities relate the kaon pseudoscalar and $\kappa$ scalar susceptibilities,
which are spacetime integrals of meson two-point correlation functions, to
the light- and strange-quark condensates.  They can therefore be used to
study the approach of these two channels to
degeneracy~\cite{GomezNicola:2020qxo}.

Tests based on two-point correlation functions have a common limitation: axial breaking may persist
in higher-point correlations even when the partner susceptibilities become
degenerate.  The flavor dependence of the 't Hooft interaction illustrates
why.  The $|Q|=1$ sectors generate the minimal 't Hooft vertex in
Eq.~\eqref{eq:thooft-vertex}, containing $2N_f$ fermion fields, or $N_f$ quark
bilinears.
For $N_f=2$, a correlator of two bilinears
can therefore couple directly to this interaction, motivating
$\Delta_{\pi\delta}$ in Eq.~\eqref{eq:delta-pidelta}.  For $N_f=3$ and $4$,
the vertex instead contains three and four bilinears.  In a chirally symmetric
phase with $N_f$ massless flavors, correlators of fewer than $N_f$ quark
bilinears are invariant under $U_A(1)$, even when axial breaking persists in
higher-point functions~\cite{Birse:1996dx}.

Carabba and Meggiolaro studied local and point-split global $2N_f$-quark
operators that are invariant under non-singlet chiral transformations but not
under $U_A(1)$.  They derived spectral relations for the global condensate
without assuming an instanton background, and separately evaluated its
high-temperature behavior in the instanton-background approximation.
For $N_f>2$ degenerate flavors, the latter calculation gives
$\Delta_{\pi\delta}\propto m^{N_f-2}$ as the common mass $m\to0$, while the
global $2N_f$-quark condensate remains nonzero in the chiral limit at finite
$T$ in this approximation and vanishes only as $T\to\infty$.  They compared
its temperature dependence with that of $\chi_t$ and proposed the global
condensate as a more accessible numerical observable than its local
counterpart~\cite{Carabba:2021xmc}.

These statements concern the limit in which all $N_f$ quark masses vanish.
Studies with $N_f\geq3$ should therefore include $2N_f$-fermion observables
or suitable free-energy derivatives.  In particular, the three-flavor chiral
limit differs from $(2+1)$-flavor QCD with $m_l\to0$ at fixed $m_s$.
With $m_s\neq0$, the strange-quark sector can supply the remaining fermion
pair, so the light $\pi$--$\delta$ difference remains a direct two-flavor
axial diagnostic.

\begin{table}[htbp]
\centering
\begin{tabular}{p{0.18\textwidth}p{0.24\textwidth}p{0.22\textwidth}p{0.24\textwidth}}
\toprule
Probe & Effective-restoration signal & Main strength & Main qualification \\
\midrule
$\chi_\pi-\chi_\delta$ & Vanishes in the thermodynamic and chiral limits & Direct, inexpensive test using non-singlet correlators & Integrated quantity; exact zero modes and order of limits matter \\
Dirac density $\rho(\lambda)$ & Infrared weight is suppressed strongly enough to make Eq.~(\ref{eq:spectral-delta}) vanish & Identifies the microscopic modes responsible & Sensitive to chiral properties of the lattice operator and to sea--valence mismatch \\
Quark-mass derivatives of $\rho$ and eigenvalue correlations & Mass dependence and correlations consistent with vanishing axial-breaking spectral integrals in the chiral limit & The derivatives probe spectral contributions to axial breaking and chiral fluctuations and are computed directly from eigenvalue correlations on a single fixed-mass ensemble. & Computing these derivatives does not require the chiral-limit assumptions; their mass dependence tests them. \\
$\chi_t$ and $Q$ cumulants & $\chi_t/m_l^2\to0$ with restored non-singlet symmetry; Ward-identity consistency & Gluonic and cosmologically relevant; tests semiclassical behavior & Net charge omits paired activity; $b_2$ alone does not test axial restoration \\
Screening correlators & $U_A(1)$ partner correlators become degenerate & Direct long-distance spatial information & Equal masses alone do not imply equal full correlators; singlets are costly \\
Higher-point functions and source derivatives & Higher-point correlators become invariant under axial rotations & Tests symmetry beyond the two-point sector & $\theta$ independence at zero mass is insufficient; source derivatives and higher-point functions are costly \\
\bottomrule
\end{tabular}
\caption{Principal diagnostics of effective $U_A(1)$ restoration.  The two-point
criteria refer to two light flavors, including $(2+1)$-flavor QCD with fixed
nonzero $m_s$; they do not establish invariance of all higher-point functions.
The continuum and thermodynamic limits precede the chiral limit at a specified
temperature.  The rows test different consequences of the symmetry and are
most informative when used together.}
\label{tab:diagnostics}
\end{table}

A controlled conclusion therefore compares several diagnostics on matched
ensembles and with the same order of limits.  Agreement tests whether the
partner channels, infrared spectrum, its mass response, topology, and
higher-point correlations describe one physical picture.  A mismatch can
instead reveal a lattice artifact or an
overinterpretation of one observable.  The systematic controls needed for
this comparison are the subject of the next section.

\FloatBarrier
\section{Lattice QCD formulations and systematic uncertainties}
\label{sec:lattice-systematics}

Lattice QCD provides a first-principles approach to the strong interaction
by replacing continuous Euclidean spacetime with a grid of spacing $a$.
For a lattice of size $N_\sigma^3\times N_\tau$, where $N_\sigma$ and $N_\tau$
are the numbers of sites in each spatial and Euclidean-time direction,
\begin{equation}
 T=\frac{1}{aN_\tau},\qquad L=aN_\sigma.
 \label{eq:lattice-temperature}
\end{equation}
The path integral becomes a statistical average that can be evaluated by
Monte Carlo sampling at zero baryon chemical potential.  In practice,
calculations begin at nonzero $a$, finite $L$, and
nonzero quark mass.  The $U_A(1)$ problem is unusually sensitive to all three
regulators because it is dominated by a small number of infrared modes and
because chiral symmetry itself is difficult to preserve on a lattice.

\subsection{Fermion discretizations and sea--valence consistency}

The Nielsen--Ninomiya theorem prevents a local, doubler-free lattice action
from retaining the naive continuum chiral symmetry~\cite{Nielsen:1981hk}.  Different formulations
choose different compromises.  Their main properties
for the present problem are summarized in Table~\ref{tab:fermions}.

\begin{table}[htbp]
\centering
\begin{tabular}{p{0.15\textwidth}p{0.28\textwidth}p{0.24\textwidth}p{0.21\textwidth}}
\toprule
Formulation & Chiral and topological advantage & Principal concern for this problem & Typical role \\
\midrule
Improved staggered (e.g. HISQ) & Computationally efficient; a remnant chiral symmetry permits light masses, large volumes, and several spacings & Four tastes become degenerate only as $a\to0$; taste breaking distorts would-be zero modes and partner channels & High-statistics thermodynamics, mass scaling, and increasingly direct eigenvalue studies \\
\addlinespace[0.45em]
Domain wall & Left- and right-handed modes live on opposite boundaries of a fifth dimension; chiral breaking can be made small & Finite fifth dimension leaves a residual mass and mode-dependent Ginsparg--Wilson violation, especially important near zero & Dynamical ensembles with good chiral symmetry and meson/spectral cross-checks \\
\addlinespace[0.45em]
Overlap & Exact Ginsparg--Wilson symmetry and an exact lattice index at nonzero $a$ & Very expensive; dynamical topology changes and determinant evaluation are difficult & Valence spectral probe or reweighted target action; some fully dynamical studies \\
\addlinespace[0.45em]
Wilson/clover & Mature algorithms and no taste multiplicity & Chiral symmetry is explicitly broken at finite $a$ and requires renormalization and improvement & Thermodynamics and scaling studies; less direct for near-zero axial diagnostics \\
\bottomrule
\end{tabular}
\caption{Fermion formulations commonly used in finite-temperature
$U_A(1)$ studies.}
\label{tab:fermions}
\end{table}

The overlap operator obeys the Ginsparg--Wilson relation
\begin{equation}
 \gamma_5D+D\gamma_5=aD\gamma_5D,
 \label{eq:gw-relation}
\end{equation}
which supports an exact lattice-modified chiral transformation and reproduces
the correct anomaly~\cite{Ginsparg:1981bj,Neuberger:1997fp,Luscher:1998pqa}.
Its index gives
an unambiguous fermionic topological charge.  This makes overlap eigenmodes an
ideal diagnostic, but generating large, light, finite-temperature ensembles
with an overlap determinant is very costly.

Domain-wall fermions approximate the same structure in five dimensions
~\cite{Kaplan:1992bt,Shamir:1993zy,Furman:1994ky}.
In the limit of infinite fifth-dimensional extent, the light
four-dimensional operator becomes overlap-like.  At finite extent, a
residual mass summarizes some chiral violation, but a single residual-mass
number does not describe every eigenmode.  Rare near-zero modes can violate
the Ginsparg--Wilson relation more strongly than bulk modes and contaminate
$\Delta_{\pi\delta}$~\cite{Cossu:2015kfa}.  Some studies correct this by
reweighting the entire domain-wall determinant to an overlap determinant.
Reweighting is exact in principle, but its statistical overlap can deteriorate
with volume.

Staggered fermions retain a remnant chiral symmetry and are much less
expensive, which enables smaller light-quark masses and continuum sequences.
One unrooted staggered field contains four fermion copies, called ``tastes'',
originating from lattice doubling.  They become degenerate as $a\to0$;
rooting the determinant is used to recover the desired number of physical
flavors in this limit.  At finite $a$, taste-breaking interactions split
their spectra.  Topological near-zero modes
approach the expected near-degenerate quartets only toward the continuum.
For instance, the HISQ action greatly reduces taste
breaking~\cite{Follana:2006rc}, but a continuum extrapolation remains essential.

For the two-flavor chiral phase transition discussed in
Section~\ref{sec:chiral-transition}, staggered fermions at nonzero lattice
spacing preserve only an $O(2)$ remnant of the continuum $O(4)$ chiral
symmetry.  Thus, if the chiral transition is continuous at fixed lattice
spacing, its asymptotic scaling is $O(2)$; $O(4)$ scaling can emerge only as
taste symmetry is restored toward the continuum limit~\cite{Ding:2015ona}.

Computing overlap eigenvalues on gauge ensembles generated with a HISQ or
domain-wall determinant combines an excellent spectral probe with affordable
sea configurations.  It also creates a \emph{mixed action}: the valence
operator used in the measurement is not the operator whose determinant
weighted the gauge fields.  At nonzero lattice spacing this is a partially
quenched, nonunitary theory.  Near-zero overlap modes may be enhanced or
suppressed differently from the sea modes, so their contribution cannot
automatically be interpreted as the spectrum of a fully dynamical overlap
theory.

There are three standard responses.  One can reweight the sea determinant to
the overlap action~\cite{Tomiya:2016jwr,Aoki:2020noz}; compare results obtained
with valence overlap modes with calculations using the sea Dirac operator
~\cite{Kaczmarek:2021ser,Kaczmarek:2023bxb}; or repeat the mixed-action
calculation at several spacings and test whether the mismatch disappears in
the continuum.  These approaches address sea--valence mismatch and cutoff
effects in different ways and should not be ranked by formulation name
alone.  Exact chiral symmetry of a valence
operator at a single lattice spacing does not by itself control sea--valence
mismatch or cutoff effects.  Conversely, calculations using the sea operator
at several spacings can quantify those effects through a continuum
extrapolation~\cite{Ding:2020xlj,Kaczmarek:2023bxb}.

\subsection{The order of limits, topology, and finite-volume control}
\label{sec:order-of-limits}

The desired critical statement contains at least three limits.  A schematic
ordering is
\begin{equation}
 \lim_{m_l\to0}\;\lim_{L\to\infty}\;\lim_{a\to0}
 \Delta_{\pi\delta}(m_l,L,a,T),
 \label{eq:limits}
\end{equation}
with the continuum and volume limits often approached through combined fits.
The thermodynamic limit must precede the chiral limit when testing spontaneous
symmetry breaking.  At finite volume, exact
non-singlet chiral symmetry cannot break spontaneously, so the condensate
vanishes at $m_l=0$; taking $m_l\to0$ first can manufacture behavior unrelated
to the infinite system.

Exact topological zero modes can strongly affect $\Delta_{\pi\delta}$ at
finite volume.  For a chirally symmetric Dirac operator, the index theorem
gives $|Q|$ topological zero modes per flavor on a configuration of charge $Q$.
Their contribution is $4\langle|Q|\rangle/(V_4m_l^2)$.  At $\theta=0$,
$\langle|Q|\rangle\leq\sqrt{\langle Q^2\rangle}=\sqrt{\chi_tV_4}$,
so this contribution vanishes as $V_4\to\infty$ at fixed nonzero $m_l$,
provided $\chi_t$ remains finite~\cite{Kanazawa:2015xna}.  Near-zero modes
whose number grows proportionally to $V_4$ can instead retain a finite
density and an axial-breaking signal in this limit.

A fixed-$Q$ calculation samples only one sector of the full $\theta=0$
ensemble and can therefore bias correlation functions at finite volume.
At fixed nonzero quark masses and fixed $Q$, a saddle-point expansion gives
corrections beginning at $O(1/V_4)$.  This expansion assumes that the
free-energy density and correlation functions are analytic near $\theta=0$,
with $\chi_tV_4\gg1$ and $|Q|\ll\chi_tV_4$~\cite{Aoki:2007ka}.
As $m_l$ decreases at fixed volume, $\chi_tV_4$ can cease to be large, so
the expansion need not control the approach to the chiral limit.

Leutwyler and Smilga made the interplay of quark mass, volume, topology, and
the low Dirac spectrum explicit.  Finite-volume partition functions in fixed
topological sectors yield spectral sum rules, while sectors of nonzero winding
number are increasingly suppressed as the quark masses decrease
~\cite{Leutwyler:1992yt}.

Within the analytic two-flavor framework of
Eq.~(\ref{eq:mass-expanded-free-energy}), summing over all topological
sectors at $\theta=0$ gives $\Delta_{\pi\delta}=8f_A$ at leading order.
Restricting the ensemble to $Q=0$ instead gives
\begin{equation}
 \left.\Delta_{\pi\delta}\right|_{Q=0}
 =8f_A\frac{I_1(x)}{I_0(x)},\qquad
 x=2V_4f_A m_l^2,
 \label{eq:fixed-topology-fA}
\end{equation}
where $I_n$ is a modified Bessel function and $x=\chi_tV_4$ at the order
retained here~\cite{Kanazawa:2015xna}.  For $x\ll1$,
$I_1/I_0\simeq x/2$, so a calculation restricted to $Q=0$ misses the full
signal.  When $m_l\to0$ is taken before $V_4\to\infty$, the finite-volume
signal comes from the rare $|Q|=1$ sectors: their combined probability,
$P(|Q|=1)\simeq2V_4f_A m_l^2$, multiplies the zero-mode contribution
$4/(V_4m_l^2)$ to give $8f_A$ in the full ensemble.  Taking $V_4\to\infty$
first at fixed $m_l>0$ instead makes the
contribution of exact zero modes vanish; nonzero modes carry the signal, while
$I_1/I_0\to1$ and the fixed-$Q$ result approaches $8f_A$.  Thus both
sequences yield the same full susceptibility within this analytic
approximation, although the contributing modes differ.  This does not
justify interchanging limits at a critical point or using
Eq.~\eqref{eq:fixed-topology-fA} as a universal correction for frozen topology.

The continuum limit is equally important.  In a fixed-$N_\tau$ temperature
scan, $T$ is varied by changing $a$, so the lattice spacing and its associated
cutoff effects change simultaneously.  A continuum result at a given physical
temperature therefore requires calculations at several $N_\tau$, with the
quark masses tuned along a line of constant physics.  In a fixed-scale scan,
$a$ and the bare parameters are held fixed while $T$ is varied through
$N_\tau$; this avoids changing the lattice spacing within one temperature
scan, but continuum control still requires comparison among several lattice
spacings.  Thus, a temperature trend observed at a single coarse lattice
spacing is not a continuum result.

Topology is difficult to sample for two opposite reasons.  Near the crossover
there can be many near-zero modes, which make light-quark inversions expensive.
At high temperature, nonzero-$Q$ configurations are rare, so an enormous
Monte Carlo history may be needed to estimate $\langle Q^2\rangle$.  As the
lattice spacing decreases, standard algorithms can also remain trapped in one
topological sector for long periods (``topological freezing'').  An apparently
small $\chi_t$ is not convincing if the Markov chain never changed $Q$.

Autocorrelations and the sampling of topological sectors should therefore be
checked carefully.  Fixed-topology methods, multicanonical sampling, and
reweighting can help, but require control of finite-volume corrections and
sampling overlap~\cite{Aoki:2007ka,Bonati:2018blm}.  Parallel tempering on
boundary conditions exchanges configurations among replicas interpolating
between periodic and locally open gauge boundaries, with observables measured
on the periodic replica.  In full QCD, combining this approach with a
multicanonical bias, removed by reweighting, addresses both topological
freezing and the rarity of nonzero charge at high
temperature~\cite{Bonanno:2024zyn}.

As a consistency check, one can compare $Q$ obtained after continuously
smoothing the gauge field
(``gradient flow'') with $Q$ inferred from the overlap-operator index, the
difference between the numbers of left- and right-handed zero modes.  The two
definitions need not agree configuration by configuration at finite $a$, but
their correlation strengthens toward the continuum
~\cite{Luscher:2010iy,Alexandrou:2017hqw}.

Volume dependence helps determine whether a near-zero peak survives the
thermodynamic limit.  At fixed $T$, $a$, and quark masses, the number of modes
under such a peak should grow in proportion to the physical four-volume
$V_4$; consequently, its integrated density per $V_4$ and its contribution to
Eq.~(\ref{eq:spectral-delta}) should approach volume-independent limits.  The
volume dependence of the peak shape and of the eigenmode localization
measures provides additional checks.

\subsection{Renormalization, scale setting, and comparable reporting}

Dirac eigenvalues renormalize in the same way as the quark mass,
$\lambda_R=Z_m\lambda$.  Correspondingly, the spectral density carries the
inverse Jacobian, $\rho_R(\lambda_R)=Z_m^{-1}\rho(\lambda)$, so that the
integrated mode number is renormalization-group invariant
~\cite{Giusti:2008vb}.  Scalar susceptibilities contain additive or
multiplicative ultraviolet pieces.
Differences such as $\Delta_{\pi\delta}$ cancel important common terms but are
not free of every normalization issue.  Studies often present combinations
such as $m_s^2\Delta_{\pi\delta}/T^4$, which are renormalization-group
invariant under the relevant multiplicative factors and dimensionless.  The
same renormalization scheme must be used when comparing spectral integrals and
direct correlators.

For a chirally symmetric lattice action, temperature subtraction removes
temperature-independent additive divergences.  Ratios of partner differences
and sums can then cancel a common multiplicative renormalization factor
~\cite{Chiu:2026upk}.  A continuum comparison also requires a fixed physical
definition of the normalization: a reference temperature held fixed in lattice
units changes as $a\to0$.  Section~\ref{sec:screening-evidence} discusses
this issue for the recent symmetry-ratio calculation.

Finally, ``the same temperature'' is not meaningful without a scale-setting
procedure and quark-mass trajectory.  Two-flavor QCD, $(2+1)$-flavor QCD
extrapolated to zero light-quark mass at a fixed physical strange-quark
mass, and $(2+1)$-flavor QCD with both light- and strange-quark masses fixed
to their physical values address different questions.  Temperatures quoted
as multiples of $T_c$ can also hide different
definitions of $T_c$.  The most comparable studies report both MeV and a
ratio $T/T_{\rm pc}$, specifying how $T_{\rm pc}$ is defined, together with
the pion or light-quark mass, $m_\pi L$, $N_\tau$, the physical volume, and
whether each observable is valence, reweighted, or fully dynamical.

These systematics are not side issues.  They explain why high-quality studies
can reach different provisional conclusions and provide the criteria for the
evidence review that follows.

\section{What lattice QCD currently shows}
\label{sec:lattice-evidence}
Several conclusions are now more secure than the final chiral-limit answer.
QCD at physical quark masses has a smooth chiral crossover near
$155$--$160$ MeV. Across this region, many susceptibility, spectral,
topological, and screening studies retain a $U_A(1)$-sensitive signal,
although the conclusion depends on the observable and correlation-function
sector~\cite{Bhattacharya:2014ara,Bazavov:2019www,Kaczmarek:2023bxb,
	Gavai:2024mcj,Ding:2026gao,Ding:2015ona}. At physical quark masses, however,
``restoration'' requires qualification: the quark mass term explicitly
breaks both non-singlet chiral symmetry and $U_A(1)$, while the anomaly
also breaks the latter. The relevant comparison therefore concerns the
rates at which the corresponding symmetry partners approach degeneracy.
More generally, axial-sensitive signals decrease as the temperature rises.
For susceptibility differences such as $\Delta_{\pi\delta}$, the infrared
contribution is controlled by the lowest Dirac modes.

The evidence is therefore best organized by the physical question each
calculation addresses, rather than by counting papers for or against
``restoration.'' A physical-mass calculation near $155$ MeV, a two-flavor
chiral extrapolation at $220$ MeV, and a continuum spectrum at fixed
physical mass are all valuable, but they do not test the same proposition.
The discussion follows three connected themes: topology and the infrared
spectrum, meson correlations and the spatial structure of Dirac modes,
and the approach to the chiral phase transition as quark masses and
flavor content change.

\subsection{Topology and the infrared Dirac spectrum}
\label{sec:topology-spectrum-evidence}

\paragraph{Topology.}
At physical quark masses, the topological susceptibility falls by orders of
magnitude from the hadronic regime into the hot plasma.  Using $(2+1+1)$-flavor lattice QCD, Borsanyi et al.
obtained a continuum-extrapolated determination of $\chi_t(T)$ over a broad
temperature range and used it to constrain the temperature-dependent axion
mass and its cosmological abundance~\cite{Borsanyi:2016ksw}.  An independent $(2+1)$-flavor
calculation with $m_\pi\simeq160$ MeV found a change in the temperature
dependence around 250 MeV, with a higher-temperature slope increasingly
compatible with DIGA~\cite{Petreczky:2016vrs}.  Agreement in slope does not
establish agreement in normalization.

Comparisons between definitions test the systematic uncertainties in $\chi_t$.
A $(2+1)$-flavor Wilson-fermion study with heavier-than-physical light quarks
used gradient flow to compare gluonic and fermionic determinations at one
lattice spacing~\cite{Taniguchi:2016tjc}.  A $(2+1+1)$-flavor maximally
twisted-mass calculation, also with heavier-than-physical light quarks, used
the relation to the disconnected chiral susceptibility in the chirally
restored regime~\cite{Burger:2018fvb}.
A staggered spectral-projector calculation in $(2+1)$-flavor QCD at physical
quark masses reduced cutoff effects and obtained continuum results.
At $T=300$ and $365$ MeV, its estimates of $\chi_t^{1/4}$ were roughly
$2$--$3$ standard deviations higher than earlier
determinations~\cite{Athenodorou:2022aay}.  Comparisons
must account for differences in flavor content and quark masses, as well
as volume and cutoff effects.

Bonanno et al. combined parallel tempering on boundary conditions with
multicanonical sampling in physical-mass $(2+1)$-flavor staggered QCD,
extending the calculation at $T\simeq570$ MeV to $a\simeq0.021$ fm.
A continuum fit including earlier ensembles substantially improved the
precision, giving $\chi_t^{1/4}=6.2(1.2)$ MeV~\cite{Bonanno:2024zyn}, compatible with an earlier
staggered determination~\cite{Borsanyi:2016ksw}.
The scale at the finest spacing was
set by extrapolation along the line of constant physics, with an additional
scale-setting uncertainty not included in the quoted error.  This progress
at 570 MeV does not settle the normalization tension at 300--400 MeV.

An earlier physical-mass $(2+1+1)$-flavor calculation by Chen, Chiu, and Hsieh
used optimal domain-wall quarks and a gluonic charge measured after Wilson
flow.  It reported a continuum determination of $\chi_t$ from a combined
temperature and $a^2$ fit to 15 ensembles spanning 155--516 MeV at three
lattice spacings, $a\simeq0.064$--0.075 fm~\cite{Chen:2022fid}.
Their continuum-extrapolated values of $\chi_t(T)$ were higher than several
staggered determinations at comparable temperatures.
The fit ansatz, the narrow range of lattice spacings, and finite-spacing
chiral effects remain relevant to this comparison.

Recent $(2+1)$-flavor calculations with chiral lattice formulations provide
further checks.  A dynamical-overlap study at $N_\tau=8$, using fixed topology
and the slab method, found substantial volume dependence; its infinite-volume
extrapolation agreed with earlier staggered results within larger
uncertainties~\cite{Fodor:2025mqi}.  Physical-point M\"obius-domain-wall
calculations at $N_\tau=12,16$, supplemented by $N_\tau=10$ at higher temperatures,
found sizable cutoff effects and difficulties sampling nonzero topological
sectors at the highest temperatures~\cite{Kanamori:2026sep}.  Neither study
yet provides a controlled continuum determination of $\chi_t$.

Higher cumulants test whether topological fluctuations approach the DIGA
prediction. Pure-Yang--Mills studies found a rapid approach of $b_2$ to
$-1/12$ above deconfinement~\cite{Bonati:2013tt,Borsanyi:2015cka}, whereas
finite-spacing studies of physical-mass $(2+1)$-flavor QCD, extending to
about $4T_{\rm pc}$, found a more gradual approach, with deviations persisting
to about $2.5T_{\rm pc}$ on the finest lattice~\cite{Bonati:2015vqz}.
A multicanonical study obtained a continuum-extrapolated value of $b_2$ compatible with
$-1/12$ near 430~MeV. When nonzero charge is rare and only $Q=0,\pm1$
contribute appreciably, $b_2$ can lie near $-1/12$ without establishing
DIGA. The observed volume scaling of higher-charge sector weights therefore
provided a stronger test~\cite{Bonati:2018blm}.

Exploratory $(2+1+1)$-flavor twisted-mass Wilson calculations with
heavier-than-physical light quarks also found high-temperature $b_2$ broadly
compatible with DIGA, without a continuum extrapolation~\cite{Burger:2017xkz}.
A subsequent physical-mass study obtained continuum-extrapolated $\chi_t$
and investigated $b_2$ and the $\theta$-dependent free energy, reporting a
rapid approach toward dilute-gas behavior above roughly 300
MeV~\cite{Kotov:2025ilm}.  However, its $b_2$ data did not support a controlled
continuum extrapolation, and the free-energy reconstruction remained
exploratory at finite lattice spacing.

  Together, these calculations
constrain topological fluctuations and their approach to DIGA; their
implications for chiral-limit $U_A(1)$ breaking require the quark-mass
dependence and comparison with axial observables through
Eq.~(\ref{eq:ward-topology}).

\paragraph{The infrared spectrum and its mass response.}

The topological susceptibility measures fluctuations of the net charge.
In a topological description, opposite-charge objects can generate near-zero
modes while their contributions to $Q$ cancel.  The full infrared spectrum
is therefore not fixed by $\langle Q^2\rangle$ alone.  Spectral measurements
connect the low-lying modes to the axial response through
Eq.~(\ref{eq:spectral-delta}).

The HotQCD collaboration's domain-wall program provided an early view with good chiral
symmetry.  On $(2+1)$-flavor ensembles with
$m_\pi\simeq200$ MeV and $N_\tau=8$, the disconnected susceptibility placed
the crossover around 165 MeV.  The $\pi$--$\delta$ difference remained nonzero
above that temperature and could be quantitatively reconstructed from an
infrared population of Dirac eigenmodes
~\cite{HotQCD:2012vvd,Buchoff:2013nra}.  The initial study and its larger-volume
spectral and susceptibility follow-up showed how mesonic susceptibilities can
indicate approximate non-singlet chiral restoration while a small infrared
spectral component still carries a sizable axial signal.

Using overlap eigenmodes as a probe of $(2+1)$-flavor HISQ ensembles, Dick
et al. found no infrared gap even at about $1.5T_{\rm pc}$, where
$T_{\rm pc}$ denotes the physical-mass crossover temperature.  Localized
near-zero modes provided the dominant contribution to the axial-breaking
measure.  At the higher temperature, the configuration-by-configuration number
of zero and near-zero modes was compatible with a Poisson distribution,
supporting a dilute topological interpretation~\cite{Dick:2015twa}.
These count statistics do not determine the connected eigenvalue correlations
as functions of $\lambda$.  The separation of a near-zero component from the
bulk spectrum motivated later studies of its mass dependence.  The use of
different sea and valence actions also requires the mixed-action checks
discussed in Section~\ref{sec:chiral-limit-evidence}.

Ding et al. calculated the first three derivatives of the Dirac spectral
density $\rho(\lambda;m_l)$ with respect to the light sea-quark mass through
connected eigenvalue correlations, as in Eq.~(\ref{eq:rho-mass-derivative})
~\cite{Ding:2020xlj}.  The calculation used $(2+1)$-flavor HISQ
ensembles at $T\simeq205$ MeV with a physical strange-quark mass, light masses
corresponding to $m_\pi\simeq55$--160 MeV, lattice spacings
$a\simeq0.12$, 0.08, and 0.06 fm, and aspect ratios between 4 and 9~\cite{Ding:2020xlj,Ding:2021jtn}.
The proceedings follow-up added susceptibility measurements at
$m_\pi\simeq55$ MeV on $N_\tau=12,16$ lattices, extending the lightest-mass
coverage beyond the original $N_\tau=8$ data~\cite{Ding:2021jtn}.

The infrared peak was traced to non-Poisson correlations among the
eigenvalues and became sharper toward the continuum.  In the infrared region,
$(\partial\rho/\partial m_l)/m_l$ and
$\partial^2\rho/\partial m_l^2$ were approximately equal and independent of $m_l$, while the third
derivative was consistent with zero.  This behavior supports an approximately
quadratic mass-dependent contribution to the infrared peak over the studied
mass range; it does not exclude a mass-independent component of $\rho$.
Whether the quadratic contribution yields a finite chiral-limit
$\Delta_{\pi\delta}$ also depends on its concentration in $\lambda$ relative
to $m_l$ in Eq.~(\ref{eq:spectral-delta}).
The observed sharpening and the spectral reconstruction of the
susceptibilities, together with the reported nonzero continuum and
chiral-extrapolated values of $\Delta_{\pi\delta}$ and $\chi_{\rm disc}$, are consistent with
an $m_l^2\delta(\lambda)$-like limiting contribution
~\cite{Ding:2020xlj,Ding:2021jtn}.  The result is particularly important
because it directly tests a nonuniform $m_l\to0$ limit rather than assuming a
smooth expansion of $\rho(\lambda;m_l)$, although the mass derivatives of
$\rho$ were not themselves extrapolated jointly to the continuum and chiral
limits. Section~\ref{sec:chiral-limit-evidence}
discusses those susceptibility extrapolations.  

At physical quark masses, Kaczmarek, Shanker, and Sharma measured the HISQ
Dirac spectrum in $(2+1)$-flavor QCD at several lattice spacings over $T=145$--176 MeV
~\cite{Kaczmarek:2023bxb}.  The near-zero peak became more pronounced as the
lattice was refined.  The continuum extrapolations in $1/N_\tau^2$ concerned
the slope and intercept of the approximately linear bulk spectrum, the
scaled lowest-eigenvalue distribution, and $\Delta_{\pi\delta}$; the peak
itself was identified through its evolution with lattice spacing.
These analyses supported a nonzero axial-sensitive susceptibility above the
crossover.  Comparisons with chiral random-matrix theory and an
interacting-instanton picture provided qualitative interpretations of the
low-lying spectral structure and its oscillations.

The same study fitted the continuum estimates of
$\Delta_{\pi\delta}/T^2$ to $A+B/T^2$, obtaining a zero of this ansatz at
$T/T_{\rm pc}=1.147(25)$ with $T_{\rm pc}=156.5$ MeV
~\cite{Kaczmarek:2023bxb}.  This is a fit-dependent estimate of the
suppression of the near-crossover contribution, extrapolated from temperatures
up to 176 MeV.  The authors distinguished this contribution from a possible
dilute-gas remnant at higher temperatures, such as the component studied at
205 MeV.  The result therefore does not establish exact restoration or
determine the chiral limit.

An independent pointwise continuum extrapolation of the staggered spectral
density at $T=230$ MeV and physical $(2+1)$-flavor masses also found a clear
infrared peak, at fixed spatial size $L\simeq 3.4$ fm and fixed physical
spectral resolution~\cite{Alexandru:2024tel}. Using a heuristic
topology-based prescription, the lowest $2|Q|$ positive staggered modes on
each configuration were identified as candidate would-be zero modes
associated with the minimal contribution required by the net topological
charge and removed. The residual infrared component remained nonzero in the
fixed-volume continuum extrapolation. At $a\simeq 0.061$ fm, where three
spatial volumes were available, the residual component was approximately
volume independent, while the contribution attributed to the would-be zero
modes decreased with increasing volume. Restricting the analysis to the
$Q=0$ sector gave a compatible result. These findings support an infrared
structure beyond the modes associated with the net topological charge,
although the mode separation is heuristic and the volume study was performed
at only one lattice spacing. Its light-quark-mass dependence remains to be
determined with continuum and thermodynamic control.

The dynamical-overlap study of Fodor et al. also examined the spectrum at
$N_\tau=8$ and physical $(2+1)$-flavor masses~\cite{Fodor:2025mqi}.
At $T=170$ MeV, a near-zero peak was visible on the largest volume,
$N_\sigma/N_\tau=5$, while smaller volumes strongly suppressed the infrared density.
This provides evidence for the peak with matched sea and valence actions
and exact lattice chiral symmetry.  The 170 MeV spectra were obtained at
fixed $Q=0$ and one temporal extent, so their volume dependence
and fixed-topology effects require further control before continuum or
chiral-limit conclusions can be drawn.

Together, these measurements constrain the temperature dependence of topology
and the infrared spectrum. The spatial structure of the modes and the
light-quark-mass extrapolations are examined in
Sections~\ref{sec:screening-evidence} and~\ref{sec:chiral-limit-evidence}.

\subsection{Meson correlations and the spatial structure of Dirac modes}
\label{sec:screening-evidence}

The spectral reconstructions above identify modes contributing to integrated
susceptibilities.  Meson correlators probe propagation, while localization
measurements characterize the spatial structure of those modes.
Recognizing the differences among screening masses, full correlators, and
integrated susceptibilities, as discussed in
Section~\ref{subsec:diagnostics_topology}, is essential when comparing
results.  Separate normalization of partner correlators removes
their relative amplitudes, so agreement of their shapes alone does not imply
equality of susceptibilities.  Each comparison requires its own quark-mass,
volume, and continuum control.

\paragraph{Susceptibilities near the physical-mass crossover.}

A physical-mass  $(2+1)$-flavor DWF study used spatial sizes of roughly 4--11 fm
and found a crossover near 155 MeV.  
Non-singlet chiral partner susceptibilities were consistent with degeneracy above about 164 MeV, 
whereas the axial-sensitive susceptibility difference remained nonzero above the crossover and became statistically 
compatible with zero only at the highest studied temperature, 196 MeV~\cite{Bhattacharya:2014ara}.  Because the calculation had one temporal
extent, it established a clear finite-spacing physical-mass pattern, but
neither a continuum extrapolation nor a chiral-limit determination.

A recent M\"obius domain-wall calculation by Gavai et al. complements these
studies by combining Ward identities with an overlap-index probe of topology
on $(2+1)$-flavor M\"obius-domain-wall ensembles at physical light and
strange quark masses.  The non-singlet crossover was
located at $158.7^{+2.6}_{-2.3}$ MeV, while the temperature dependence of the
topological susceptibility and its comparison with Ward-identity-related
observables indicated persistent axial-sensitive effects up to the highest
studied temperature, 186 MeV, on $N_\tau=8$ lattices
~\cite{Gavai:2024mcj}.  
This reinforces the separation between the non-singlet crossover and axial-sensitive effects at physical masses.
The next necessary steps are additional $N_\tau$, volumes, and light
masses.

Chiu and Hsieh studied ratios constructed from temperature-subtracted,
quark-connected flavor-nonsinglet meson susceptibilities in
physical-point $N_f=2+1+1$ QCD with optimal domain-wall fermions at
three lattice spacings. Their global fit in temperature and lattice
spacing yielded continuum-extrapolated ratios statistically compatible
with zero at $T=164$~MeV~\cite{Chiu:2026sxy}. In the finite-temperature
scan, this temperature was simulated only at the coarsest spacing;
the two finer scans reached down to 179 and 192~MeV. The continuum
estimate at 164~MeV therefore depends on the assumed temperature
dependence below the ranges sampled at the finer spacings.
The reference prescription fixes $aT_r=1/4$, rather than a common
physical $T_r$, so $T_r$ increases as $a$ decreases.
If the renormalized partner difference at $T_r$ is negligible,
its reference contribution to the numerator is negligible, but the
reference susceptibility sum remains in the
denominator~\cite{Chiu:2026upk}. The continuum extrapolation thus
also varies the physical reference temperature entering the
normalization. A comparison at fixed physical $T_r$, or a separate
determination of the renormalized partner differences with a
specified renormalization and subtraction prescription, would test
the interpretation of the small extrapolated ratios as degeneracy
of the corresponding susceptibilities.

A strange-sector test follows from Ward identities relating the $K$ and
$\kappa$ susceptibilities to the light- and strange-quark chiral condensates.
Using existing lattice condensate data, G\'omez Nicola et al. reconstructed
these susceptibilities and found that $\chi_S^\kappa$ develops a maximum
above the crossover before approaching $\chi_P^K$ at higher temperatures
~\cite{GomezNicola:2020qxo}.  In the light-quark chiral limit, restored
non-singlet chiral symmetry already enforces this equality through the Ward
identities.  The comparison tests the consistency of the restoration pattern
but does not isolate $U_A(1)$ breaking independently.

\paragraph{Screening correlators.}
Screening correlators provide information without an explicit low-mode
projection.  An early $(2+1)$-flavor study with the improved p4 staggered
action found vector--axial-vector degeneracy near the crossover, whereas the
pseudoscalar--scalar correlators approached degeneracy only above about
$1.3T_c$~\cite{Cheng:2010fe}.  It supplied an early indication that
non-singlet chiral restoration and the suppression of two-point $U_A(1)$
breaking need not occur at the same temperature.

Continuum-extrapolated $(2+1)$-flavor HISQ results with near-physical pion masses find that
non-singlet chiral partners become approximately degenerate
around the crossover, whereas $U_A(1)$ partners approach one another more
slowly with increasing temperature~\cite{Bazavov:2019www}.  They use the
physical pion mass at 140--172 MeV and
$m_\pi\simeq160$ MeV up to about 2.5 GeV, with continuum extrapolations up to
about 1 GeV; quark-mass effects are small at high temperature.
This establishes a usful physical-mass baseline, but  do not determine the two-flavor chiral limit.

An earlier $N_f=2$ study with $O(a)$-improved Wilson fermions at
$N_\tau=16$ and pion masses of about 200--540 MeV found the scalar--pseudoscalar
screening-mass splitting near the crossover reduced by at least a factor of
three and statistically consistent with zero~\cite{Brandt:2016daq}.
Because this was a single-$N_\tau$ study at relatively heavy masses, without
a continuum extrapolation, its implication for the continuum chiral
transition remained open.  It
nevertheless showed early that 
screening masses could indicate strong axial suppression even where 
integrated-susceptibility studies with different setups retained sizable signals.

Rohrhofer et al. compared the full set of $J=0$ and $J=1$ spatial isovector
correlators in two-flavor domain-wall QCD~\cite{Rohrhofer:2019qwq}.
Each correlator was normalized at the first nonzero spatial separation.
At about 220 MeV, the normalized scalar and pseudoscalar correlators agreed
within errors on the finer lattice at that temperature, while a coarser
ensemble retained a visible splitting.
This dependence on lattice parameters is important when interpreting partner
degeneracy.  The study also followed axial-related tensor channels, extending
the tests beyond the scalar--pseudoscalar pair and beyond fitted screening
masses alone~\cite{Glozman:2025twe}.

A recent calculation by the JLQCD collaboration follows screening masses in $N_f=2$ M\"obius
domain-wall QCD from 147 to 330 MeV, with an estimated $T_{pc}\simeq165$ MeV at the physical 
light-quark mass of this two-flavor theory and
a residual mass of about 0.14 MeV~\cite{Aoki:2025mue}.  The temperature scan
uses one lattice spacing, $a\simeq0.075$ fm, and has no continuum
extrapolation.  At the smallest
light-quark masses, vector and axial-vector screening masses become degenerate
near $T_{pc}$.  
The pseudoscalar--scalar and tensor--axial-tensor comparisons are
less conclusive directly at $T_{pc}$, while the corresponding
splittings are small and broadly compatible with approximate
degeneracy around 190 MeV and above.

The optimal-domain-wall program of Chiu and collaborators began with an
exploratory two-flavor comparison of scalar and pseudoscalar susceptibilities
with the low Dirac spectrum~\cite{Chiu:2013wwa}.  Later work studied complete
sets of connected spatial meson correlators at physical $N_f=2+1+1$ masses,
first in the light sector and then across strange and charm channels
~\cite{Chiu:2023hnm,Chiu:2024jyz}.  The later study used unnormalized
correlators to retain the relative amplitudes discarded by the earlier
normalization.  An $N_f=2+1+1+1$ calculation extended the comparison to
bottom-quark channels with physical strange, charm, and bottom masses but
$m_\pi\sim700$ MeV, at one lattice spacing, $a\simeq0.0303$ fm, and one
spatial volume~\cite{Chiu:2024bqx}.  These connected-correlator studies
reported a flavor-hierarchical approach to partner degeneracy.  The
continuum topological susceptibility from the same program
~\cite{Chen:2022fid}, discussed in Section~\ref{sec:topology-spectrum-evidence},
provides complementary information.  Together, these finite-mass results do
not fix the chiral-limit fate of all $U_A(1)$-sensitive correlators.

A controlled high-temperature benchmark is supplied by Dalla Brida et al.,
who computed non-singlet screening masses with three massless quarks at
12 temperatures from about 1 to 160 GeV.  Three or four lattice spacings at
each temperature allowed continuum extrapolations with uncertainties of a
few parts per thousand~\cite{DallaBrida:2021ddx}.  Scalar--pseudoscalar and
vector--axial-vector masses were degenerate, while a vector--pseudoscalar
splitting remained resolved even at the highest temperature.  For three
massless flavors, these two-point degeneracies do not determine axial
breaking in higher-point functions, as discussed in
Section~\ref{subsec:diagnostics_topology}.

\paragraph{Temporal non-singlet correlators.}
The FASTSUM collaboration studied temporal correlators on anisotropic
Wilson--clover ensembles with $N_f=2+1$ quarks.  Each channel was normalized
at the temporal midpoint, and restricted time sums excluded short distances
to reduce Wilson-term artifacts.  The authors' criterion for degeneracy of
these normalized pseudoscalar and scalar observables yielded
$T_{U_A(1)}=319(22)$ MeV on their finest temporal lattice, with
$m_\pi\simeq380$ MeV and $T_{pc}=182(1)$ MeV~\cite{Aarts:2026kpq}.
The result concerns normalized temporal correlators at finite lattice
spacing and heavy light-quark masses; it does not establish equality of the
full susceptibilities.  A continuum physical-mass determination remains open.

\paragraph{The thermal singlet channel.}
The $\eta'$ offers a direct return to the hadronic motivation for the anomaly.
Kotov, Lombardo, and Trunin extracted its finite-temperature mass from
Euclidean-time correlators of the gradient-flowed topological charge density in
$N_f=2+1+1$ twisted-mass Wilson QCD, with pion masses of approximately
210 and 370 MeV~\cite{Kotov:2019dby}.  Their results were compatible with a modest dip near
the crossover and an increase at higher temperature.  This exploratory
singlet-channel measurement complements nonsinglet partner comparisons, but
does not determine their chiral limit.  A thermal singlet mass also depends
on mixing and on the correlator used to extract it; the zero-temperature
Witten--Veneziano relation cannot simply be applied with $\chi_t(T)$ inserted.

\paragraph{Localization.}
Early indications of Dirac-mode localization came from
instanton-liquid models~\cite{Garcia-Garcia:2005azc} and
finite-temperature staggered studies
~\cite{Garcia-Garcia:2006vlk,Gavai:2008xe}.
High-temperature calculations in quenched $SU(2)$ gauge theory
identified localized low modes separated from extended bulk modes
by a mobility edge~\cite{Kovacs:2010wx}; this pattern was
subsequently observed in physical-mass $(2+1)$-flavor QCD
~\cite{Kovacs:2012zq}.
In the latter theory, finite-size scaling and multifractal
analyses support the three-dimensional unitary Anderson
universality class~\cite{Giordano:2013taa,Ujfalusi:2015nha},
and a renormalized mobility edge has been extrapolated to
the continuum at 230 MeV~\cite{Bonanno:2023mzj}.

In the sea/islands picture, local Polyakov-loop fluctuations
trap low modes within an ordered background. Evidence from
quenched $SU(2)$ simulations~\cite{Bruckmann:2011cc} was
followed by support from Dirac--Anderson models and
M\"obius domain-wall studies
~\cite{Giordano:2016cjs,Cossu:2016scb}.
A recent physical-mass $(2+1)$-flavor staggered study places
the localization onset near 150--160 MeV, within the chiral
crossover region, although this onset has not been
extrapolated to the continuum~\cite{Giordano:2026bqj}.
Its proximity to the crossover does not by itself establish
$U_A(1)$ restoration.

The bulk mobility edge does not settle the behavior of the
modes closest to zero. Overlap studies in pure-$SU(3)$ gauge
theory, without dynamical sea quarks, found increasing spatial
extent toward $\lambda=0$, motivating a proposed critical edge
at the origin~\cite{Alexandru:2021pap,Alexandru:2021xoi}.
Physical-mass $(2+1)$-flavor calculations found
temperature-dependent volume scaling of near-zero mode
support~\cite{Meng:2023nxf}, while overlap probes on
domain-wall ensembles showed infrared level statistics
between Poisson and random-matrix behavior
~\cite{Pandey:2024goi,Shanker:2026mtu}.
Overlap valence studies with twisted-mass Wilson sea quarks
also exposed limitations of mobility-edge estimates based
solely on the inflection point of the relative mode volume
~\cite{Kehr:2023wrs,Kehr:2025nth,Kehr:2026wzn}.
These results do not yet establish the near-zero structure
in the continuum and chiral limits.

At nonzero quark mass, localized near-zero modes can carry
a substantial $\Delta_{\pi\delta}$~\cite{Dick:2015twa}.
Their contribution toward the chiral limit depends on both
their spectral weight and its mass response.
A theoretical constraint applies when non-singlet chiral
symmetry is restored and scalar and pseudoscalar
susceptibilities are finite. Assuming ordinary Dirac spectral
densities and connected two-eigenvalue correlations at
nonzero mass, with correlation bounds uniform in mass,
a nonzero chiral-limit $\Delta_{\pi\delta}$ is incompatible
with modes localized all the way down to the origin below
a mobility edge bounded away from zero
~\cite{Giordano:2024jnc,Giordano:2025fcr}.
A delocalized near-zero band or a mobility edge approaching
zero sufficiently rapidly can avoid this restriction;
critical behavior confined to $\lambda=0$ alone cannot
under these assumptions.
Joint mass and volume studies of Dirac eigenmodes, their
correlations, and $\Delta_{\pi\delta}$ offer a concrete way to
identify which modes sustain axial breaking and how their
contribution changes toward the chiral limit.

\subsection{Quark-mass dependence, flavor, and the chiral phase transition}
\label{sec:chiral-limit-evidence}

The physical-mass patterns and continuum spectra discussed above do not
by themselves determine the chiral limit. This requires comparing the mass
dependence of axial observables and following the chiral phase transition of the same
theory. The two-light-flavor limit at fixed strange-quark mass must also be
distinguished from the limit in which three or more flavors become massless
together.

\paragraph{Light-quark-mass dependence of axial observables.}
Building on an earlier dynamical-overlap study at fixed topology~\cite{Cossu:2013uua}, 
the JLQCD collaboration subsequently demonstrated how lattice chiral artifacts can distort the inferred axial signal~\cite{Tomiya:2016jwr}.
 Its two-flavor M\"obius domain-wall ensembles probed temperatures around $T\simeq200$ MeV at
$a\simeq0.08$--0.12 fm, reaching bare light-quark masses of a few MeV.
Despite the small residual chiral-symmetry breaking, sizable
mode-by-mode violations of the Ginsparg--Wilson relation in near-zero
modes could dominate $\Delta_{\pi\delta}$.  Using overlap fermions only
in the valence sector introduced a further distortion, since unmatched
overlap zero modes were not suppressed by the sea determinant.  Overlap
observables controlled the valence chiral violation, while determinant
reweighting corrected the sea--valence mismatch.  The resulting
zero-mode-subtracted $\Delta_{\pi\delta}$ was strongly
suppressed at the lightest masses, and its linear chiral extrapolation
was consistent with zero. The authors interpreted this as consistent with vanishing
toward the chiral limit, while acknowledging that their linear fit could
miss higher-order mass dependence~\cite{Tomiya:2016jwr}.

In a later study, the JLQCD collaboration examined this picture on finer lattices~\cite{Aoki:2020noz}, with $a\simeq0.074$ fm,
finding similarly strong suppression toward small quark masses over the wider temperature range $T\simeq190$--330 MeV.
A dedicated volume comparison at $T\simeq220$ MeV and checks against
the earlier coarser ensembles broadened the assessment of systematic effects.
The study also examined meson and baryon correlators on the underlying
M\"obius domain-wall ensembles; these did not use the overlap reweighting
applied to the spectra and susceptibilities.
The light-mass lever arm for a chiral
extrapolation nevertheless remained limited: only the lightest mass, where simulated,
lay below the estimated physical-mass reference, which relied on leading-order
chiral perturbation theory and a single zero-temperature pion measurement.
The observed mass dependence supported the authors' interpretation of
vanishing axial breaking in the chiral limit. The study, however, did not perform a continuum extrapolation, so this did not yet constitute a continuum-extrapolated chiral-limit determination.

Related
$(2+1)$-flavor calculations reweighted to
overlap fermions examined the Dirac spectrum, $\Delta_{\pi\delta}$, and
$\chi_t$ at four temperatures between 136 and 204 MeV, with light-quark
masses around and below the physical value
~\cite{Aoki:2022ebi,JLQCD:2024xey}.  The lightest point,
$am_l=0.001$, was obtained by mass reweighting from ensembles generated
at $am_l=0.002$, at a fixed lattice spacing of about 0.08 fm.
At 153--204 MeV,
$\Delta_{\pi\delta}$ was strongly suppressed at the lightest mass, while it
remained nonzero at 136 MeV.  The displayed susceptibility was not zero-mode
or ultraviolet subtracted, unlike in some earlier analyses.  These results
remain preliminary and do not constitute a continuum chiral-limit
determination~\cite{JLQCD:2024xey}.

The reweighting comparisons establish that residual chiral violation and
sea--valence mismatch can qualitatively alter the inferred near-zero signal.
Relating the suppression observed in the later two-flavor study to
criticality requires approaching the transition of that same theory and mass
trajectory.  The physical $(2+1)$-flavor crossover is not its temperature
reference.  Reweighting also becomes statistically harder as the volume grows.

Using both $\Delta_{\pi\delta}$ and $\chi_{\rm disc}$, the HISQ study at 205 MeV
discussed in Section~\ref{sec:topology-spectrum-evidence} obtained a nonzero
anomaly-sensitive two-point susceptibility after combined continuum and
chiral extrapolations. Together with the infrared peak described there,
the nonzero extrapolated values of $\Delta_{\pi\delta}$ and $\chi_{\rm disc}$
show that $U_A(1)$ breaking remains appreciable at about 205 MeV
~\cite{Ding:2020xlj}.  For comparison, a continuum-extrapolated $(2+1)$-flavor calculation in the
two-light-flavor chiral limit, $m_l\to0$ at fixed physical strange-quark mass,
obtained $T_c=132^{+3}_{-6}$ MeV~\cite{HotQCD:2019xnw}. Thus 205 MeV is
well above the estimated chiral-limit transition temperature, and the
persistence of the signal suggests that axial breaking in these two-point
functions may become more pronounced closer to the transition.
This temperature dependence cannot, however, be established from a
single-temperature calculation.   Repeating the analysis at several
temperatures near the mass-dependent pseudocritical line would connect it
directly to the universality question.

Kaczmarek, Mazur, and Sharma followed a different route: $(2+1)$-flavor HISQ
ensembles at physical strange-quark mass and $m_s/m_l=27,40,80$, corresponding
to pion masses of about 135, 110, and 80 MeV, were probed using 100--200
overlap eigenvalues per configuration.  The two heavier masses used
$32^3\times8$ lattices, while the lightest used $56^3\times8$ lattices.
The spatial extent increased at the lightest mass, without a volume scan
at matched mass and temperature.
The renormalized infrared spectrum did not develop a gap.
At $T\simeq1.05T_{pc}(m_l)$, where $T_{pc}(m_l)$ denotes the mass-dependent
pseudocritical temperature, the authors fitted
$m_l^2\Delta_{\pi\delta}/T^4$ at three masses as a polynomial in $m_l^2/T^2$
with zero constant term.  The coefficient of the term linear in
$m_l^2/T^2$ implied a finite, nonzero
$\Delta_{\pi\delta}/T^2$ as $m_l\to0$~\cite{Kaczmarek:2021ser}.
This $N_\tau=8$ study provides evidence for persistent axial breaking just
above the mass-dependent crossover, using valence masses matched through a
renormalized condensate combination. Extending the analysis to finer lattices
and several spatial volumes, while monitoring residual mixed-action effects,
would test whether the nonzero chiral-limit result survives the continuum
and thermodynamic limits.

\paragraph{Approaching the chiral phase transition.}

Building on the JLQCD collaboration's susceptibility decomposition at $T\gtrsim190$ MeV
~\cite{Aoki:2021qws}, a proceedings study extended the same two-flavor
M\"obius-domain-wall and overlap-reweighted program down to 147 MeV
~\cite{Aoki:2024uvl}.  Mass reweighting extended its coverage to about
one fifth of the physical light-quark mass.  The axial contribution remained important
near the pseudocritical region, although its dominance became harder to
establish in the noisier low-temperature, light-mass data.  Preliminary peak
estimates gave $T_{pc}\simeq165$ MeV at the physical-light-quark-mass point of
this $N_f=2$ theory and $T_c\simeq$153 MeV in the chiral limit.  The single-spacing
analysis does not yet establish continuum or critical scaling.

An exploratory continuation of the 205 MeV analysis examined spectral mass
derivatives at 137--176 MeV with HISQ fermions on $N_\tau=8$ lattices and
$m_\pi\simeq110$ MeV~\cite{Ding:2021gdy}.  In this region,
$m_l^{-1}\partial\rho/\partial m_l$ no longer agrees with
$\partial^2\rho/\partial m_l^2$ as it did at 205 MeV; the second derivative
even becomes negative at some lower temperatures.  This indicates a change
in the spectral mass response near the pseudocritical region.  The study
uses one light sea mass and one temporal extent.  Extending this
analysis to several masses and spacings would test the relation between this
mass response and the critical spectral correlations discussed in
Section~\ref{subsec:spectral-correlations}.

The spectral calculation discussed in
Section~\ref{subsec:spectral-correlations} connects chiral cumulants to
weighted spectral correlations with $O(2)$ scaling at nonzero lattice
spacing~\cite{Ding:2023oxy}.  Comparing this scaling with the mass and
temperature dependence of $\Delta_{\pi\delta}$ would test how the same
infrared modes contribute to the axial response; the cumulant result alone
does not settle that question.

Recent $(2+1)$-flavor HISQ studies have also tested critical scaling using a subtracted
chiral order parameter, $M=M_l-H\chi_l$, where $H=m_l/m_s$ and
$\chi_l=\partial M_l/\partial H$ for the normalized light-quark condensate
$M_l$.  Ratios of $M$ at different light masses allow $T_c$ and the exponent
$\delta$ to be determined without fixing $\delta$ in advance, provided
subleading corrections are sufficiently small.  Preliminary $N_\tau=8$
results were compatible with $O(2)$ behavior at the smaller masses, while
ratios involving the physical mass showed sizable corrections
~\cite{Mitra:2024mke}.  A subsequent finite-volume analysis extended the
coverage to $H=1/240$, or $m_\pi\simeq45$ MeV, at physical strange-quark
mass.  The lightest masses agreed with $O(2)$ finite-size scaling, with
additional large-volume data and finer lattices still needed
~\cite{Mitra:2025hsk}.  These studies improve control of the scaling region,
but do not yet distinguish the candidate continuum universality classes.

An independent comparison comes from $(2+1+1)$-flavor maximally twisted
Wilson fermions.  Kotov, Lombardo, and Trunin found behavior compatible with
the $O(4)$ equation of state at the physical pion mass and obtained a
chiral-extrapolated transition temperature of $134^{+6}_{-4}$ MeV
~\cite{Kotov:2021rah}.  The temperature estimate was relatively insensitive
to the assumed universality class.  These results support the $O(4)$
interpretation, while studies at lighter quark masses and a controlled continuum
extrapolation are needed to distinguish it from alternative critical scenarios.

\paragraph{Flavor dependence and the continuum phase boundary.}
As discussed in Section~\ref{sec:chiral-transition}, the phase boundary
constrains the anomalous interactions relevant near the chiral phase transition.
For three flavors, the leading determinant interaction is cubic and can
drive a first-order transition~\cite{Pisarski:1983ms,Pisarski:2024esv}.
Within determinant-based effective descriptions, reducing the leading
anomalous coupling can shrink the first-order region.  A shrinkage induced
by changing the lattice spacing or action, however, is not a measurement
of a weaker physical anomalous coupling.  The phase boundary also depends
on the remaining interactions and their infrared flow.
Moreover, disappearance of the first-order region does not by itself imply
restoration of $U(1)_A$.  In mean-field studies, the first-order region
can disappear when the leading cubic anomalous coupling vanishes while
couplings to higher-order axial-breaking operators remain nonzero
~\cite{Giacosa:2024orp}.  Direct axial-symmetry tests, including
higher-point observables, are therefore needed.

Lattice studies have not yet established the continuum phase boundary.  Early coarse-lattice
staggered and Wilson calculations found first-order signals for three light
flavors~\cite{Brown:1990ev,Iwasaki:1996zt}; 
a subsequent staggered study characterized the endpoint's
three-dimensional Ising behavior using unimproved actions
and obtained a substantially lower critical pseudoscalar mass
with improved gauge and p4 staggered fermion actions,
showing significant action dependence at
$N_\tau=4$~\cite{Karsch:2001nf}.
With unimproved staggered quarks, de Forcrand, Philipsen, and collaborators
mapped the $N_f=3$ and $(2+1)$-flavor critical lines, found a reduced
three-flavor critical mass with RHMC rather than the inexact $R$ algorithm,
and observed substantial cutoff dependence between $N_\tau=4$ and
preliminary $N_\tau=6$ results
~\cite{deForcrand:2003vyj,deForcrand:2006pv,deForcrand:2007rq}.
These studies supported the conventional Columbia-plot picture at nonzero
lattice spacing, but not its continuum boundary.

Stout-smeared staggered calculations followed two different mass
trajectories.  Endr\H{o}di et al. used a Symanzik gauge action and approached
the chiral corner along a $(2+1)$-flavor trajectory with nearly physical
$m_l/m_s$, obtaining small critical-mass upper bounds on $N_\tau=4,6$
lattices~\cite{Endrodi:2007gc}.
These do not locate the endpoint on the three-flavor-degenerate axis.
On that axis, a study with a Wilson plaquette gauge action and two-step
stout smearing found that the critical bare mass decreased rapidly with
smearing on $N_\tau=4$, but more mildly on $N_\tau=6$; small volumes and
limited statistics precluded a continuum conclusion~\cite{Varnhorst:2015lea}.

With HISQ fermions on the degenerate axis, no first-order signal was found
down to a pseudoscalar mass near 80 MeV on $N_\tau=6$, with a scaling-based
upper bound $m_\pi^c\lesssim50$ MeV~\cite{Bazavov:2017xul}; an $N_\tau=8$
analysis was compatible with a second-order chiral transition and $O(2)$
scaling at nonzero lattice spacing~\cite{Dini:2021hug}.
The Wilson--clover sequence exhibits particularly large cutoff effects:
an initial continuum extrapolation from $N_\tau=6,8$ gave a critical
pseudoscalar mass near 300 MeV, whereas adding $N_\tau=10$ and then
$N_\tau=12$ lowered the estimated upper bound to about 170 and 110 MeV,
respectively~\cite{Jin:2014hea,Jin:2017jjp,Kuramashi:2020meg}.
Unrooted $N_f=4$ staggered results on $N_\tau=4,6,8,10$ showed analogous
softening, excluding rooting as the sole explanation; comparison with
Wilson results also cautioned against naive continuum extrapolation
~\cite{deForcrand:2017cgb}.
The cutoff trend is consequently more informative than any single critical
mass.

Tricritical scaling provides an alternative route to the chiral limit.
A tricritical point marks where the transition in the massless theory
changes between first and second order as another parameter is varied.
Near this point, the boundary between first-order and crossover regions
follows a characteristic power-law dependence on the quark mass, allowing
calculations at nonzero masses to constrain the chiral limit.
Studies using unimproved staggered fermions have applied this approach by
varying the imaginary chemical potential, the number of flavors, and the
lattice spacing.
Bonati et al. traced this boundary by varying the imaginary chemical
potential in two-flavor QCD; their extrapolation indicated a first-order
chiral transition at zero chemical potential on coarse $N_\tau=4$
lattices~\cite{Bonati:2014kpa}.
Cuteri, Philipsen, and Sciarra subsequently mapped the critical boundary
for degenerate $N_f=2$--8 quarks on $N_\tau=4,6,8$
lattices~\cite{Cuteri:2021ikv}.
For $2\leq N_f\leq6$, their scaling analysis favored the critical quark mass
vanishing at a nonzero lattice spacing.
This would mark a tricritical point at which the first-order region
disappears and the massless transition becomes second order on finer
lattices.
D'Ambrosio et al. found evidence for the same behavior at fixed imaginary
chemical potential~\cite{DAmbrosio:2025ldv}.
These results support a second-order continuum chiral transition, an
interpretation that further studies on finer lattices and with other
fermion discretizations can test.
A recent many-flavor study reports this behavior through $N_f=7$.
For $N_f=8$, the first-order thermal transition instead meets a bulk
transition at finite quark mass, obscuring the approach to the chiral
limit~\cite{Klinger:2026pbe}.

Most recently, a three-flavor M\"obius domain-wall study found crossover-like
finite-size behavior at the investigated pseudocritical masses on
$N_\tau=6,8,12$, all at fixed $a\simeq0.136$ fm~\cite{Zhang:2026zpa}.
At the lowest temperature, about 121 MeV, the pseudocritical renormalized
quark mass was about 3.5--3.7 MeV in the $\overline{\rm MS}$ scheme at
2 GeV, depending on the susceptibility used to locate it.
The available volumes at this temperature do not exclude a weak first-order
transition.
The temporal extents sample different temperatures, not a continuum
sequence, and the study does not determine the continuum chiral limit.

The above results together favor a shrinking first-order phase transition region, while its continuum extent
remains unresolved.  Phase boundaries and direct axial tests, including
higher-point observables (Section~\ref{subsec:diagnostics_topology}), provide
additional constraints.  

\section{What the lattice QCD results imply}
\label{sec:questions}

The studies in Section~\ref{sec:lattice-evidence} have identified
near-zero Dirac modes that contribute to axial susceptibility
differences and measured how the Dirac spectrum responds to the sea-quark
mass. Topological measurements and spatial meson correlators provide
further tests of the underlying dynamics. These advances make
microscopic descriptions testable beyond reproducing a single
susceptibility: they can be confronted with the mass dependence
and spatial distribution of the axial response. The following
discussion examines what these results imply for the origin of
axial breaking and its role in the chiral phase transition.

Understanding axial breaking requires more than finding a temperature
at which a selected signal becomes statistically compatible with zero.
Exact vanishing in a specified limit is a well-defined theoretical
statement; numerical compatibility with zero depends on statistical
sensitivity and systematic control. Nor does a small susceptibility
difference at one temperature and quark mass establish that axial
breaking is irrelevant to critical behavior. At a continuous phase
transition, this requires understanding how the axial response scales
as the chiral correlation length grows; no universal numerical
threshold for the susceptibility difference settles that question.
The central questions are how axial breaking is generated and how
its effects change with distance, temperature, and quark mass.
A effective restoration temperature alone leaves these dynamical connections
unexplained. A calculation that distinguishes microscopic mechanisms
or connects spectral structure to meson propagation helps answer
these questions even before the limiting behavior of a selected
signal is settled.

\subsection{Topology and the microscopic origin of axial breaking}
\label{sec:axial-mechanisms}

Topology is a relevant microscopic ingredient, but $\chi_t$ alone
cannot identify the mechanism responsible for axial breaking.
In descriptions based on topological objects, weaker net-charge
fluctuations can result from fewer objects or from stronger
correlations between opposite charges. Local topological-density
correlations, higher cumulants of $Q$, and the associated fermion
modes help distinguish these possibilities. The dilute instanton
gas supplies a high-temperature reference; its success there does
not establish how topological objects interact near the
crossover~\cite{Gross:1980br,Petreczky:2016vrs}.

Interacting instantons and anti-instantons, neutral molecular
configurations, and calorons with constituent dyons offer candidate
descriptions~\cite{Shuryak:1993ee,Schafer:1994nv,
	Kraan:1998pm,Lee:1998bb}. Lattice QCD studies by Larsen et al.\
found zero-mode profiles and a dependence on the valence-quark
temporal boundary condition consistent with constituent
dyons~\cite{Larsen:2018crg,Larsen:2019sdi}.
To explain axial breaking, such a description must also reproduce
$\Delta_{\pi\delta}$, its quark-mass dependence, and the spatial
difference $G_\pi(z)-G_\delta(z)$, while satisfying the Ward identities.
Correlated instanton--anti-instanton pairs have zero net topological
charge but can modify the Dirac modes and meson correlators.
Their induced interactions can preserve $U_A(1)$~\cite{Schafer:1994nv},
so observing these pairs does not establish that axial-partner
differences survive the chiral limit.

The spectral mass-response measurements reviewed in
Section~\ref{sec:topology-spectrum-evidence} provide a test beyond
reproducing the average Dirac spectrum at one quark mass. Microscopic
descriptions that produce similar infrared peaks can predict
different changes as the sea-quark mass decreases.
The susceptibility $\Delta_{\pi\delta}$ weights the average
eigenvalue density, whereas the disconnected chiral susceptibility
$\chi_{\rm disc}$ weights its derivative with respect to the light
sea-quark mass. Through the fermion determinant, this derivative
is related to connected eigenvalue
correlations~\cite{HotQCD:2012vvd,Kanazawa:2015xna,Ding:2020xlj}.
Where $\chi_\pi=\chi_\sigma$ holds,
$\chi_{\rm disc}=\Delta_{\pi\delta}$ constrains their spectral
integrals, without imposing a pointwise relation between the density
and its derivative. Comparing both observables tests whether the
proposed near-zero dynamics accounts for axial breaking and
condensate fluctuations together.

\subsection{Magnitude and spatial range of axial breaking}
\label{sec:axial-range}
The magnitude and spatial range of axial breaking characterize
different aspects of the same correlation functions.
For the $\pi$ and $\delta$ channels, $\Delta_{\pi\delta}$ integrates
$G_\pi(z)-G_\delta(z)$ over spatial separation, whereas screening
masses characterize the large-distance decay of each correlator.
Equal screening masses therefore do not require
$\Delta_{\pi\delta}$ to vanish: the correlators can still differ
in amplitude or at shorter separations.
Likewise, a susceptibility difference compatible with zero within
numerical precision does not establish equality of the
long-distance correlators. Susceptibility differences and
screening-mass splittings can therefore suggest different
apparent restoration temperatures.

Following $G_\pi(z)-G_\delta(z)$ as temperature changes would show
where the susceptibility difference is being reduced. The partner
difference might decrease across a broad range of separations or
become concentrated at shorter distances. Retaining the relative
amplitudes of the correlators is essential to distinguish these
changes, with quark masses and lattice artifacts controlled as
discussed in Section~\ref{sec:screening-evidence}.

Dirac eigenvectors provide spatial information absent from the average
eigenvalue density. At nonzero quark mass, localized near-zero modes
can contribute substantially to
$\Delta_{\pi\delta}$~\cite{Dick:2015twa}.
Their contribution to $G_\pi(z)-G_\delta(z)$ depends on
eigenfunction overlaps, including terms involving different modes.
A population that dominates the integrated susceptibility need
not dominate the correlator difference at large separation.
Identifying which modes contribute at which distances would
therefore distinguish microscopic descriptions that reproduce
the same integrated axial signal.

The chiral-limit constraints in
Section~\ref{sec:screening-evidence} also make localization part
of this interpretation. Under the assumptions stated there,
including finite scalar and pseudoscalar susceptibilities,
a nonzero chiral-limit $\Delta_{\pi\delta}$ is incompatible with
modes remaining localized down to the origin below a mobility
edge bounded away from zero~\cite{Giordano:2024jnc,Giordano:2025fcr}.
A delocalized near-zero component or a mobility edge approaching
zero can avoid this restriction. These assumptions do not
directly cover a critical point with divergent susceptibilities.

A partner difference concentrated at short separations does not
by itself establish that axial breaking is irrelevant to criticality.
Local anomalous interactions can still change which meson modes
become critical; their influence must be assessed on the scale
of the growing chiral correlation length.

\subsection{Axial breaking at the continuum chiral phase transition}
\label{sec:axial-criticality}

The chiral-limit studies in Section~\ref{sec:chiral-limit-evidence}
leave unresolved how anomalous interactions affect the order of
the transition and, if it is continuous, its critical behavior.
A fixed-temperature chiral extrapolation above the transition
establishes the axial response in that regime. Connecting this
response to criticality requires following axial-partner
correlations as the chiral correlation length grows.

For two massless light flavors approaching a continuous transition
from the chirally symmetric phase, the $\sigma$ and pion correlation
lengths diverge. The axial question concerns the light-quark
pseudoscalar singlet $\eta$ and scalar isovectors $\delta^a$.
If their screening masses stay nonzero, their correlation lengths
remain finite, allowing a critical order-parameter theory containing
the $(\sigma,\pi^a)$ modes and $O(4)$ universality.
If the $\eta$ and $\delta$ correlation lengths also diverge,
these modes must be included. Effective axial symmetry in the
critical theory would additionally require the $\pi$--$\delta$
and $\sigma$--$\eta$ correlators to share their leading critical
forms, and the leading higher-point critical correlations to
respect $U_A(1)$. The effective-theory analysis in
Section~\ref{sec:chiral-transition} describes how anomalous
interactions can select between these possibilities
~\cite{Pisarski:1983ms,Pelissetto:2013hqa,Pisarski:2024esv}.

Axial-partner correlators and chiral cumulants measured on the
same ensembles, for example along the mass-dependent
pseudocritical line, would connect these symmetry tests to the
developing critical fluctuations. Comparing the mass scaling
of $\Delta_{\pi\delta}$ with $\chi_{\rm disc}$ and higher
cumulants would also test how the eigenvalue correlations
governing condensate fluctuations contribute to the axial
response~\cite{Ding:2023oxy}. The finite-mass spectral and
scaling results reviewed in Section~\ref{sec:chiral-limit-evidence}
provide starting points for this comparison.

Flavor dependence requires a further qualification.
For three or more simultaneously massless flavors, two-point
degeneracy in the chirally symmetric phase can coexist with
axial breaking in higher-point functions. Measurements of
$2N_f$-quark correlations or suitable free-energy derivatives
are then needed~\cite{Birse:1996dx,Carabba:2021xmc}.
The shrinking first-order regions reviewed in
Section~\ref{sec:chiral-limit-evidence} constrain the phase boundary,
but do not by themselves measure a weaker anomalous interaction
or establish axial restoration. Combining phase-boundary studies
with direct axial tests would clarify that connection
~\cite{Pisarski:2024esv,Cuteri:2021ikv}.
This differs from $m_l\to0$ at fixed strange-quark mass, where
light-sector two-point tests retain sensitivity to axial breaking.

For a continuous phase transition, the decisive result would combine
chiral scaling with explicit axial-partner tests under controlled
continuum, thermodynamic, and chiral limits.
It would determine which meson channels become critical and
whether their leading correlations respect $U_A(1)$.
This would establish how microscopic anomalous interactions
shape the symmetries and collective fluctuations of the
chiral phase transition.

\section{Concluding perspective: how does hot QCD remember the anomaly?}
\label{sec:perspective}

Hot QCD offers a way to follow an exact quantum effect into collective
behavior. Its anomalous Ward identity remains fixed while the thermal
ensemble reorganizes. The progress reviewed here makes this connection increasingly
testable: microscopic descriptions can be confronted with both
the response of the Dirac eigenvalue density to the sea-quark
mass and the meson correlations of the same ensemble.

For two light flavors, one concrete target is to trace the near-zero
contribution to $\Delta_{\pi\delta}$ into the spatial correlator difference
$G_\pi(z)-G_\delta(z)$. At fixed temperature and nonzero quark mass,
decomposing the quark propagators into Dirac eigenmodes would resolve
how near-zero modes contribute to this difference at each spatial separation.
The spatial and volume dependences of this contribution would distinguish
descriptions that agree on the susceptibility but predict different
propagation. Repeating the comparison at lighter sea-quark masses and
measuring mass derivatives of the Dirac eigenvalue density would test
how this connection changes toward the chiral limit. Extending this
analysis to three or more quark flavors becoming massless together
requires higher-point quark correlations, since two-point degeneracy
in the chirally symmetric phase can coexist with axial breaking.
Such comparisons would advance the dynamical explanation before the
order and critical behavior of the chiral phase transition are settled.

The same topological density enters the physics of the vacuum $\eta'$ mass~\cite{Witten:1979vv,Veneziano:1979ec},
thermal singlet correlations and the axion potential~\cite{Lombardo:2020bvn}, and real-time chirality production~\cite{Kharzeev:2024zzm}.
These observables probe different correlations and scales: an equilibrium
axial susceptibility cannot by itself determine a thermal singlet mass
or a real-time transition rate. Testing a microscopic description across
these scales would clarify how the same gauge dynamics produces such
different physical consequences.

Understanding these connections would reveal how microscopic quantum
structure shapes the fields and symmetries of hot matter at long
distances. The anomaly itself is exact; how hot QCD remembers it remains
a dynamical question.

\begin{ack}[Acknowledgments]
	
The author is supported by the National Natural Science Foundation of China under Grants No. 12325508, No. 12293064 and No. 12293060.
\end{ack}

\bibliographystyle{Numbered-Style}

\enlargethispage{18pt}
\bibliography{reference}

\end{document}